\documentclass[useAMS,usenatbib]{mn2e}

\usepackage{graphicx}

\title{Bayesian Cluster Finder: \\
Clusters in the CFHTLS Archive Research Survey}
\author[B. Ascaso, D. Wittman \& N. Ben\'itez]{B. Ascaso$^{1}$\thanks{E-mail:
ascaso@physics.ucdavis.edu};  and D. Wittman$^{1}$ and N. Ben\'itez$^{2}$.\\
$^{1}$Physics Dpt. One Shields Av. Davis, CA 95616, USA\\
$^{2}$Instituto de Astrof\'isica de Andaluc\'ia (CSIC), Apdo. 3044, 18008 Granada, Spain}
\begin{document}

\date{Accepted 31st October 2011. Received 10th June 2011}

\maketitle

\label{firstpage}

\begin{abstract}
The detection of galaxy clusters in present and future surveys enables measuring mass-to-light ratios, clustering properties, galaxy cluster abundances and therefore, constraining cosmological parameters. We present a new technique for detecting galaxy clusters, which is based on the Matched Filter Algorithm from a Bayesian point of view. The method is able to determine the position, redshift and richness of the cluster through the maximization of a filter depending on galaxy luminosity, density and photometric redshift combined with a galaxy cluster prior that accounts for color-magnitude relations and BCG-redshift relation. We tested the algorithm through realistic mock galaxy catalogs, revealing that the detections are 100\% complete and 80\% pure for clusters up to z $<$1.2 and richer than $\Lambda_{CL}>$20 (Abell Richness $\sim$0, M$\sim$4$\times10^{14} M_{\odot}$). The completeness and purity remains approximately the same if we do not include the prior information, implying that this method is able to detect galaxy cluster with and without a well defined red sequence. We applied the algorithm to the CFHTLS Archive Research Survey (CARS) data, recovering similar detections as previously published using the same or deeper data plus additional clusters which appear to be real.
\end{abstract}


\section{Introduction}

Galaxy clusters are the largest virialized structures in the Universe. They are valuable laboratories for investigating environmental effects on galaxy evolution \citep{treu03,moran07} as well as the properties of their dark matter content. In addition, the evolution of their global properties through time and distance are of great importance to constrain the cosmological parameters of the Universe.  Therefore, it is important to define samples of clusters which span a wide range of redshift and cluster mass with a minimum of false positives while remaining as complete as possible.

Achieving this goal requires very good wide-field survey data as well as very good algorithms.  In the past decade, survey data have vastly improved both in terms of quantity (SDSS, DLS, CFHTLS...) and quality (depth, resolution, wavelength coverage, and uniformity).  Additional surveys recently begun or under construction including Pan-STARRS, KIDS, DES, HSC and LSST will deliver even more and better data.  In this context, it is important to develop cluster-selection algorithms which take maximum advantage of the large amount of information available.

A number of methods to detect clusters have been developed based on cluster X-ray emission \citep{rosati02}, weak lensing \citep{tyson90,wittman01,wittman03} and the Sunyaev-Zeldovich effect (SZ:  \cite{carlstrom02,ascaso07,menanteau09}).  Furthermore, the cluster detection methods based on optical data have provided a large dataset of clusters. They take advantage of the larger number of bands, better photometric redshift quality or improvement in depth of the new surveys. We summarize below the main optical cluster detection methods in three main groups based on the use of different optical characteristics.

The first group of methods is based on the geometric distribution of the galaxies. They usually create density maps based on the spatial distribution of the sources. The candidate clusters are identified as overdensities over the mean background. These methods include the Counts in Cells method \citep{couch91,lidman96}, the Percolation Algorithms \citep{dalton97,botzler04}, the Voronoi Tessellation algorithm \citep{ramella01,kim02,lopes04} or the Friends of Friends  Algorithm \citep{ramella02,botzler04,vanbreukelen09}. They have the advantage that they do not bin data or make any assumption prior to the detection process. Thus, non-regular structures can be detected. However, these methods usually have large false detection rates \citep{couch91,lidman96}, or do not go as deep as other methods  \citep{kim02,lopes04}. Some variants have been introduced lately taking into account photometric redshifts or colors \citep{kim02}, magnitudes \citep{ramella01,lopes04} or redshifts \citep{way10} in the Voronoi Tessellation Method to minimize the background and foreground contamination.

The second group takes into account the presence of the red sequence of the clusters and the fact that the early type population, including the brightest cluster galaxy (BCG), of the clusters have predictable colors and in some cases, magnitudes \citep{ascaso11}. Some algorithms based on these properties are the Cut-and-Enhance Algorithm \citep{goto02}, the Cluster Red Sequence Method \citep{gladders00,lopezcruz04,gladders05}, MaxBCG \citep{hansen05,koester07} and the C4 cluster-finding algorithm \citep{miller05}. Some of these algorithms have been successfully applied to surveys such as SDSS or the Red Sequence Cluster Survey (RCS). Generally, those methods generate a set of color slices based on a particular model and select subsets of galaxies to belong to each slice. Then, peaks in the density maps are selected. Those algorithms work well in sufficiently deep photometric catalogs and provides robust photometric z estimates. However, the presence of the red sequence in all kind of clusters is still a matter of debate. Although many studies have proved it to exist up to redshift $\sim$ 1 \citep{mei06,ascaso08,andreon08,mei09}, others \citep{donahue02} argued that the red sequence could be a characteristic only for virialized massive clusters while less massive clusters may not show it even at moderate redshifts. Thus, this method could be biased towards the high end of the mass function.

A higher redshift version of this algorithm has been implemented with SpARCS (Spitzer Adaptation of the Red-Sequence Cluster Survey; \citealt{wilson08}), which samples the 4000$\AA$ break by analyzing Spitzer 3.6 micron and z band data. So far,  some clusters have been confirmed spectroscopically \citep{wilson09,muzzin09,demarco10}. However, the main concern about this method is the possible bias towards clusters showing the red sequence feature.

Finally, the third group includes those methods where different cluster characteristics such as the luminosity or density  are taken into account to model the clusters.  In general, the density and luminosity distributions of galaxies are first modeled for different redshift slices and consequently probability density maps are generated. Then, clusters are selected as the peaks of these maps and their redshift and richness estimated. The Matched Filter \citep{postman96,postman02} has been applied to detect clusters in different surveys such as the Palomar Distant Cluster Survey. Some later modifications include the Adaptive  Matched Filter \citep{kepner99}, the Hybrid Matched Filter \citep{kim02}, the Simple Smoothing Kernels \citep{shectman85}, the Adaptive Kernel method  \citep{gal00,gal03} or the 3D-Matched Filter \citep{milkeraitis10}. They have been applied to NoSOCS \citep{gal00}, CFHTLS \citep{olsen07,grove09,milkeraitis10} and SDSS \citep{dong08,szabo10}. These methods usually recover a high completeness rate and small contamination. However, the results might be model dependent. A slightly different algorithm that can be included in this group is the Surface Brightness (SB) Enhancements \citep{zaritsky97,zaritsky02}, which consists of detecting the localized cumulative SB enhancement due to unresolved light from galaxies in distant clusters. 

Another important variation of this technique at high redshift has been introduced by \cite{eisenhardt08}. They used 3D overdensities based on the photo-z distribution. In their simulations, they found that only 10\% of the detections are spurious. They identified 106 galaxy clusters at z$>$1 in 7.25 square degrees in the Spitzer Infrared Array Camera (IRAC) Shallow Survey. To date, they have confirmed 12 of them spectroscopically within a redshift range $1<z<1.41$.

In addition, new ways of detecting methods at higher redshift have been developed for detecting galaxy clusters around radio galaxies at z$\sim$ 1.5-2 \citep{galametz09,chiaberge10}. 

The motivation of this work is to take advantage of all the characteristics of the present methods by modeling each cluster property. There might be some cluster properties that do not show up in certain ranges of redshift or mass, for example the CMR, but we still want to detect a cluster if one of these properties is not present. To accomplish this, we have designed a Bayesian cluster finder where each galaxy is assigned a Bayesian probability that there is a cluster centered in that galaxy at a given redshift.  This method takes into account cluster properties such as the cluster luminosity function, density profiles and photometric redshift distribution to generate a likelihood probability as a matched filter technique variation. In addition, we have included cluster properties such as the presence of a red sequence and the BCG magnitude - redshift relation as a prior term. Although these are properties that have been well observed, it is not clear that they exist and are present in every kind of cluster. With this formalism, we can easily turn on or off the prior term and as a result, be able to detect any kind of cluster. The uniqueness of this method comes from the inclusion of different sets of cluster properties.

We have tested the algorithm extensively on simulations. We have derived the completeness and purity rates of the detections and we  have studied the improvement or dependence of our detections on the introduction of such a prior.  Moreover, we have also tested the algorithm on real data. We have applied the algorithm to CFHTLS-Archive-Research Survey, (CARS, \cite{erben09}) and compared our results with those obtained by  \cite{olsen07},  \cite{adami10} and X-ray spectroscopically confirmed detections in the same fields. We find the results to be in fairly good agreement, providing support to our results.

The structure of the paper is as follows. In section 2, we describe the method we have implemented. Section 3 is devoted to simulations to study the completeness, purity and false positive rate of the results. Section 4 shows the cluster detections in CARS data \citep{erben09} and the comparisons with other works. In Section 5, we include a summary and discussion of the results. Where appropriate, we have used the $\Lambda_{CDM}$ cosmology $H_0$=71 km s$^{-1}$ Mpc$^{-1}$, $\Omega_M$ =0.23, $\Omega_L$=0.73 throughout this paper.

\section{The method}

We consider that a cluster in an $(X,Y)$ position, can be defined by $N_g$, $R_c$ and $z_c$, that is the number of galaxies in a given aperture ('richness'), the cluster radius and a redshift. 

The set of galaxies in the field or in a cluster can be described by their coordinates $(x_i,y_i)$, photometric redshift $z_i$ (including color information), magnitudes $m_i$ and spectral types $t_i$ estimated from a photometric redshift algorithm. 

The probability for the existence of a cluster centered on the position $(X,Y)$, given a dataset $D$ and a priori information $I$ is

\begin{equation}
p(X,Y,N_g,R_c,z_c | D, I)
\end{equation}

Finding this probability is proportional (applying Bayes' Theorem) to 

\begin{eqnarray}
\lefteqn{p(X,Y,N_g,R_c,z_c | D, I)  \nonumber }\\ 
& \propto p(X,Y,N_g,R_c,z_c | I) p(D| X,Y,N_g,R_c,z_c)
 \end{eqnarray}
 
where the first term
\begin{equation}p(X,Y,N_g,R_c,z_c | I)\end{equation}
is the prior, or the information which is not implicit in the data, such as the distribution of the cluster properties, cluster redshift and spatial cluster distribution, and the second term is the likelihood
\begin{equation}p(D| X,Y,N_g,R_c,z_c)\end{equation}
which is the probability of observing galaxy data D if there is a cluster in $(X,Y)$ with $z_c$, $N_g$ and $R_c$ (also known as the likelihood).

The likelihood can be rewritten as

\begin{eqnarray}
L(X,Y,N_g,R_c,z_c)=p(D| X,Y,N_g,R_c,z_c) = \nonumber\\ 
 \prod_{i=1}^{N}p(x_i,y_i,z_i,m_i,t_i | X,Y,N_g,R_c,z_c)
\end{eqnarray}

or similarly, as
\begin{eqnarray}
\lefteqn{ \ln L(X,Y,N_g,R_c,z_c)  =  \nonumber }\\ 
& \sum_{i=1}^{N} \ln p(x_i,y_i,z_i,m_i,t_i | X,Y,N_g,R_c,z_c)  
\end{eqnarray}

That is, the product of the likelihood for each galaxy, where the likelihood is the probability of observing a galaxy $(x_i,y_i)$, with a given redshift, magnitude and spectral type, if there is a cluster at the $(X,Y)$ position with $N_g$, $R_c$ and $z_c$.

\subsection{Likelihood}

The likelihood or predicted density probability ($n$) for a cluster can be described by 
     
 \begin{eqnarray}
\lefteqn{\ln L(X,Y,N_g,R_c,z_c)= \nonumber}\\
& \sum_{i} n (x_i,y_i,z_i,m_i,t_i| X, Y,N_g,R_c,z_c)=\\
& \sum_i P(x_i,y_i | X,Y,R_c,N_g,z_c))L(m_i,z_i | z_c) p(z_i,t_i | z_c)\nonumber
\end{eqnarray}
where $P(r)$ is the cluster radial profile.  The sum is made over all the galaxies within a radius of $r_{cut}$ from the central one and within the limits of reliability of the photo-z. 

We use the Plummer profile to model $P(r)$  \citep{postman96}. 

\begin{equation} 
P(r)=\left\{ \begin{array}{rl}
\frac{1}{\sqrt{1+(r/r_c)^2}}-\frac{1}{\sqrt{1+(r_{cut}/r_c)^2}}; &\mbox{ if $r<r_{cut}$}\\
0 & \mbox{ if $r>r_{cut}$}
\end{array} \right.
\label{eq:plummer}\end{equation}
where $r$ is calculated as

\begin{equation}r(x_i,y_i| X,Y,z_c)= D_A(z_c)\theta_{iC}(x_i,y_i|X,Y)\end{equation}
where $D_A(z_c)$ is the angular diameter distance at the cluster's redshift $z_c$.

We choose the core and cut radius as $r_c=0.150$ Mpc and $r_{cut}=10r_c$ respectively. 

As far as the luminosity filter is concerned, we have adopted a total luminosity profile described by  

\begin{equation}L(m_i,z_i | z_c)=b(m_i)+\phi(m_i,z_i | z_c)\end{equation}
where $b(m_i)$ is the number of background galaxies and $\phi(L)$ is the standard Schechter luminosity function  \citep{schechter76}

\begin{equation}\phi(L_i,z_i | z_c)=0.4 (\ln 10) n^* (L_i/L^*)^{1+ \alpha} exp(-L_i/L^*)\end{equation}
being $L_i=L_i(m_i,z_i)$ the luminosity for each galaxy within $r_C$ and $L^*=L^*(z_c)$ the characteristic luminosity correspondent to the luminosity function of a cluster at a given redshift $z_c$.  We have applied k-corrections to the magnitudes by using the K-correct package from \cite{blanton07}. We selected the values of $M^*(0)$=-20.44 and $\alpha=-1.05$  from \cite{blanton03} in $r$ band for local clusters, where $M^*$ is the magnitude associated to $L^*$. We also applied the following evolutionary correction to $M^*$,

\begin{equation} M^*(z)=M^*(z=0)-z \end{equation}
extracted from \cite{postman02}, which is consistent with observations \citep{postman01,nakata01,brown07}.
 
We estimate the background galaxy counts $b(m)$ in any given dataset in two iterative steps. We first detect galaxy clusters by using the background counts extracted from GOODS in R band from \cite{harsono09} as an initial estimate. Then, we obtain the survey's background galaxy distribution by masking out these detections from the survey. Finally, we run the detection algorithm again with the right background distribution.

The uncertainties on galaxy redshifts can be included in the probability with the following factor

\begin{equation}p(z_i | z_c) = \int_{z_c-\sigma_c}^{z_c+\sigma_c} p_{PDF}(z_i | z_c) dz \end{equation} 
where $p_{PDF}$ is the redshift probability distribution function for each galaxy obtained by any photometric redshift code like BPZ \citep{benitez00}. The limits of the integral are set to account for the spread of the redshifts of the galaxy cluster and their photometric errors and the fact that the estimated redshift could be shifted from the actual redshift value by $\Delta z/2$, where $\Delta z$ refers to the difference between two consecutive redshift slices. We have chosen $\sigma_c=(1+z_c)0.06$.  The actual value of  $\sigma_c$ is set from the data quality of the survey that we are simulating.

In case the  $p_{PDF}$ for each galaxy is not available, we can model $p(z_i | z_c)$ by assuming a Gaussian distribution for the redshift distribution \citep{dong08} with the same dispersion as before, $\sigma_c$.

\begin{equation}p(z_i | z_c)=exp[-(z_i-z_c)^2/2\sigma^2_{c}]/\sqrt{2\pi\sigma_{c}}\end{equation} 

\subsection{Prior}

The prior probabilities for the cluster distributions refer to already known properties of clusters that can help enhancing the likelihood. We decompose them into two different parts. 

The first term models the probability of a spatial grouping of galaxies having the expected red sequence of a cluster at its redshift. This probability can be quantified as the departure of the data from the expected red sequence of clusters at this redshift. 

\begin{equation} p(col_i)=\exp[-(col_i-col(z_c))/2\sigma^2_{col_i}]\end{equation} 
where $col_i$ refers to the slope of the color-magnitude relation for the galaxies within a 1.5 Mpc and within a photometric redshift range of $z_c \pm \Delta z/2 \pm 0.01(1+z)$ and $col(z_c)$ is the expected slope at a particular redshift slice $z_c$.

\begin{figure}
\centering
\includegraphics[width=1.0\hsize]{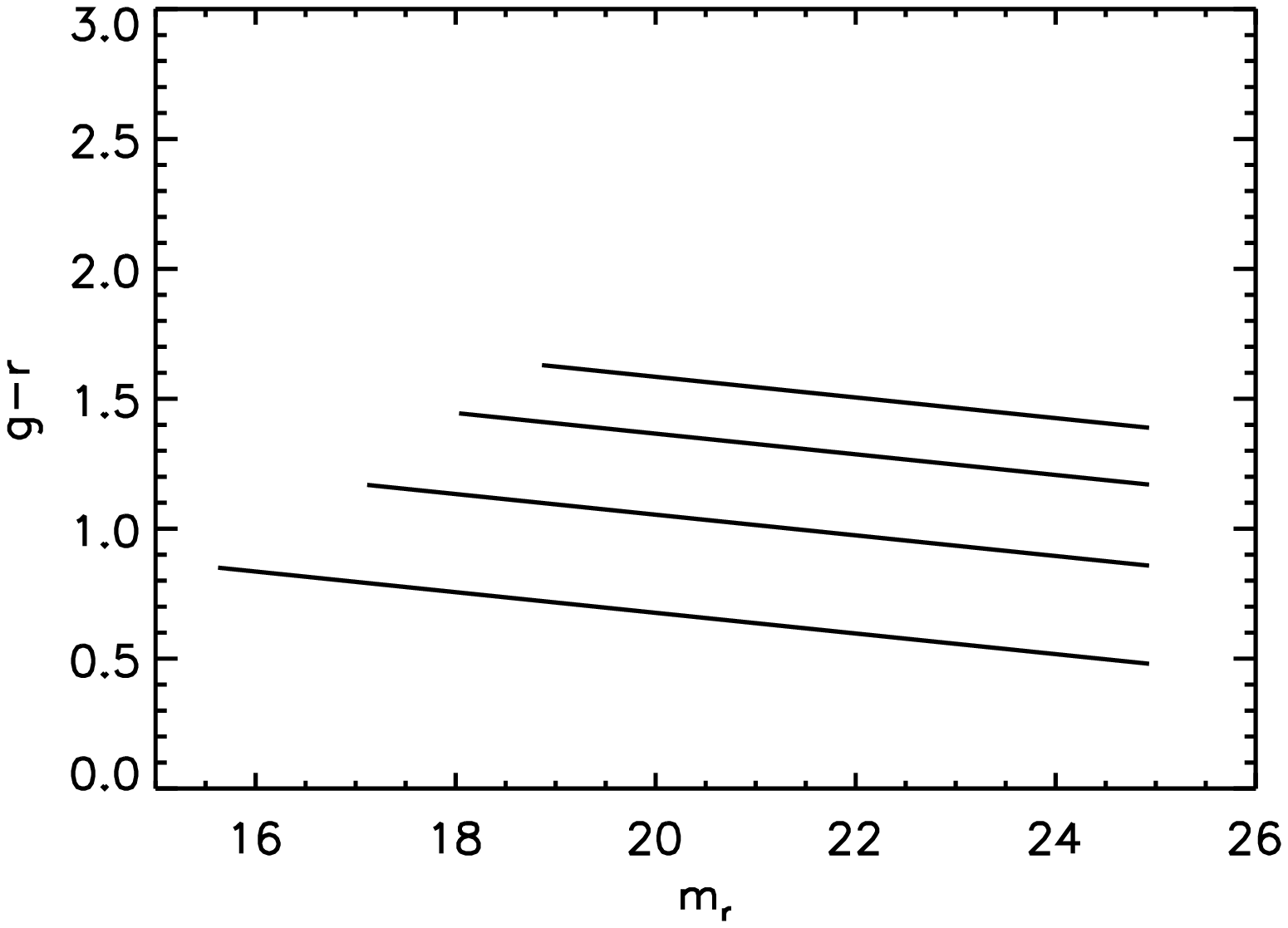} 
\includegraphics[width=1.0\hsize]{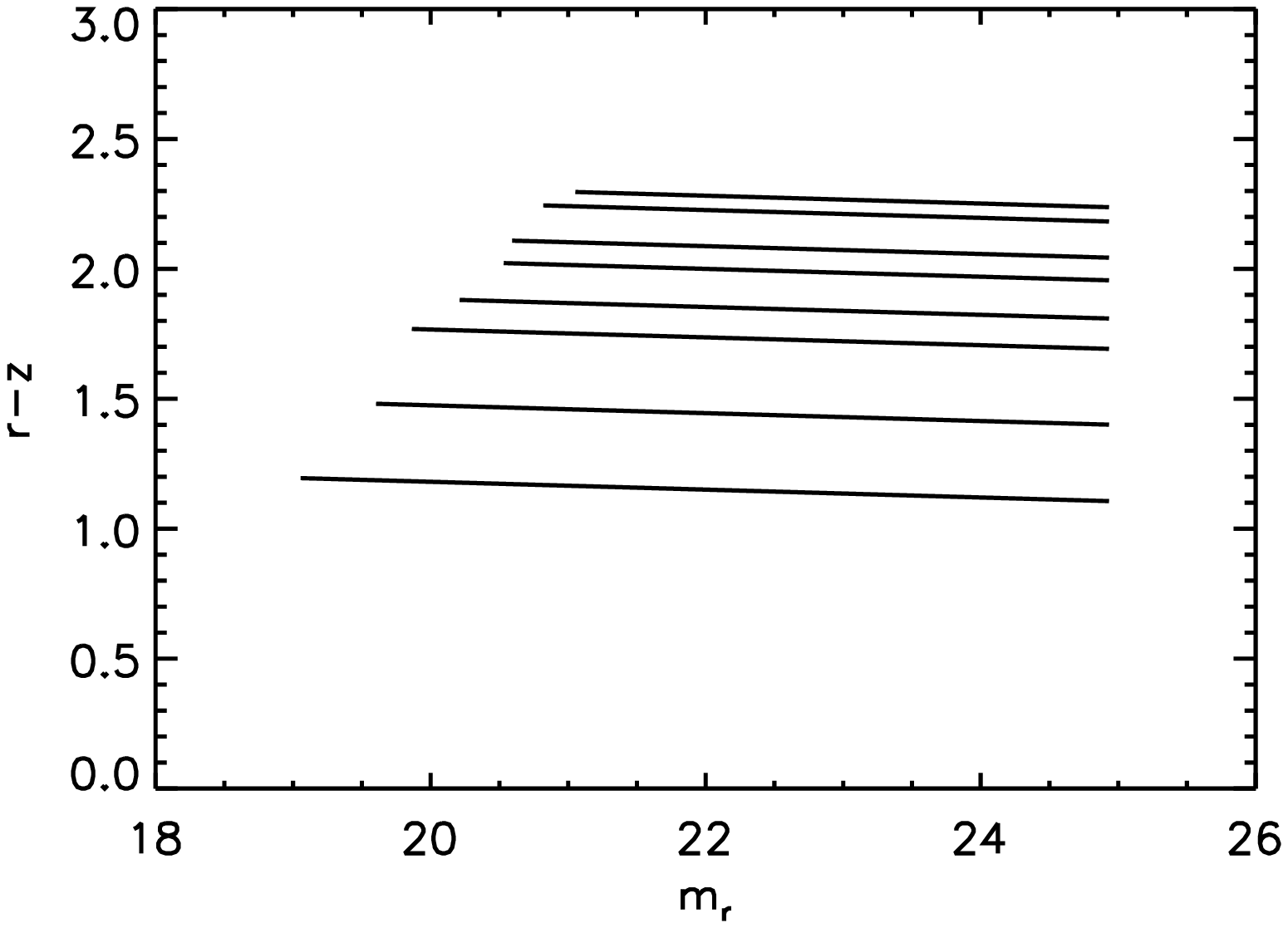} 
\caption{g-r(upper plot) and r-z (bottom plot) synthetic galaxy clusters color-magnitude for  redshift slices ranging from z=0.1 to z=0.4 (g-r) and  from z=0.5 to z=1.2 (r-z). 
}
\label{fig:CMrelat}
\end{figure}

We estimate synthetic color-magnitude relations computing the expected colors from a set of template spectra of E/S0, Sbc, Scd, and Irr by \cite{coleman80} and of starbursting galaxies SB3, SB2 by \cite{kinney96} and adding a fixed slope obtained from an individual well-characterized galaxy clusters in the survey being modeled. We selected colors that sample the 4000$\AA$ break  \citep{gladders00,gladders05,wilson06} for each redshift slice. For instance, we chose $g-r$ color for $z_s\le$ 0.3 and $r-z$ color for $z_s>$ 0.3 for the CARS simulation. The expected $g-r$ and $r-z$ CMR for each redshift slice are shown in Figure \ref{fig:CMrelat}.

The second term takes into account the fact that each cluster has a brightest cluster galaxy (BCG) and that there is a tight relation between the apparent magnitude of the BCG and the redshift of this BCG and therefore the cluster redshift \citep{gladders00,koester07}. We obtain an empirical BCG magnitude- redshift relation by fitting the 13823 BCGs sample from SDSS \citep{koester07b} and extrapolating it to higher redshift.

We model this term in the prior with an additional Gaussian term, 

\begin{equation} p(m_{BCG}(z))=\frac{\exp[-(m_i-m_{BCG}(z))^2/2\sigma^2_{m_{BCG}}]}{\sqrt{2\pi}\sigma_{m_{BCG}}}\end{equation} 
where $m_i$ refers to the magnitude of the brightest galaxy within 1.5 Mpc aperture from each galaxy, $m_{BCG}(z)$ is the expected magnitude of the BCG at the expected cluster redshift and $\sigma_{m_{BCG}}$ refers to the magnitude-redshift dispersion obtained from the empirical fit.  

\subsection{Cluster determination}

Computing the probabilities for each galaxy in the survey is only the first step. In this section, we describe how we select, merge and purify the list of cluster candidates.

\begin{figure}
\centering
\includegraphics[clip,angle=90,width=1.0\hsize]{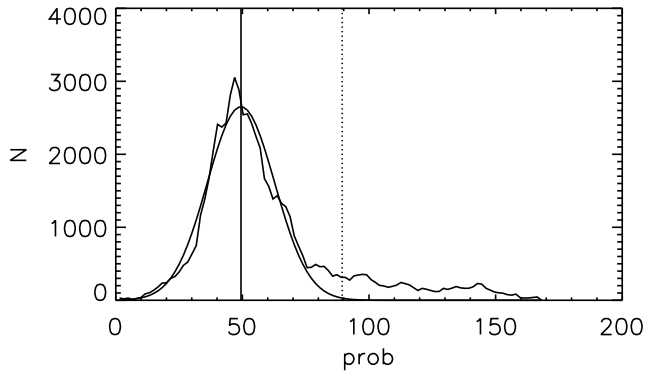} 
\caption{Background probability for a redshift slice of z=0.3. The jagged curve is the probability histogram, and the smooth curve is a Gaussian  fit to it. The solid vertical line refers to the background probability while the dotted line indicates the 3$\sigma$ threshold above which we select the galaxies belonging to clusters.}
\label{fig:gauss}
\end{figure}

First, we determine the probability threshold for cluster candidates as follows. For each redshift slice, we first compute the background probability level and its dispersion $\sigma$ by fitting a Gaussian to the whole probability distribution (for the lowest redshift slice, z=0.1, we use a log-normal to better fit the asymmetric probability distribution). The background level and $\sigma$ are set to the first and second moment of this Gaussian. Then, we select only those galaxies that have probability larger than 3$\sigma$ above background. In Figure \ref{fig:gauss}, we illustrate this with the probability distribution of a redshift slice of  z=0.3.

After that, for each galaxy we assign the redshift slice where its probability achieves a maximum over all the redshift slices.  Then, for each redshift slice, we will obtain only the galaxies with maximum probability over the threshold. 

Next, we make density maps using the selected galaxies for each redshift slice and we define galaxy clusters from the  density peaks. For each redshift slice, $z_c$, we select the maximum peak of probability and set the initial center of the candidate cluster on it. We assign $z_c$ as the redshift for the detection. We find the boundary of the cluster by finding where the radial probability profile is nearly flat, specifically $|\frac{dp}{dr}|<$ 0.1. All the galaxies included within this boundary are used for the following calculation. We calculate an additional redshift estimate from the photometric redshift distribution of the galaxies within 1.5 Mpc from this center, $z_{est}$. We fit a Gaussian to the redshift distribution of galaxies in the clusters and we adopt the peak value as $z_{est}$. 

The cluster richness is estimated with the $\Lambda_{CL}$ parameter. It is the effective number of $L^*$ galaxies in the cluster and within 1.5Mpc. Simulations in  \cite{postman02} showed that this parameter is empirically related to $N_A$, the Abell richness, that is the number of galaxies brighter than $m_3 + 2$ in the central $1.5 Mpc$, $m_3$ being  the third brightest member in the cluster. This richness is redshift-dependent, since it depends on the depth of the data. Therefore we have applied a correction based on the fraction of luminosity that we miss due to the completeness of the survey. This correction can be written in the following way:

\begin{equation} 
\Lambda_{CL}(z)=\Lambda_{CL,0}(z)\frac{\int_{m_{bright}}^{m_{faint}}\phi(m)dm}{\int_{m_{bright}}^{m_{co}}\phi(m)dm}
\end{equation}
where $m_{co}$ is the detection limit of the survey, and $\Lambda_{CL,0}$ and $\Lambda_{CL}$ are the corrected and uncorrected richness respectively (see \citealt{schuecker98}).

We run this process iteratively to the next maximum peak that has not been previously assigned to a cluster until there are no more galaxies above the threshold not assigned a cluster. With this preliminary list of clusters, we recenter the detection on the BCG candidate by iteratively looking for the brightest cluster member within 1.5 Mpc and within a redshift distance from the resdhift slice of $\frac{\Delta z}{2}(1+z)$. From this new center, we recalculate $z_{est}$ as before, keeping the boundary fixed.

Finally, we filter the preliminary list of cluster detections. First, we check that the redshift which maximizes the probability is consistent with the photometric redshift of the member galaxies. Thus, we selected as real clusters those that meet the following requirement, 

\begin{equation} 
|  z_{c} - z_{est} | \le \Delta z (1+z_ {c})
\label{eq:odd}
\end{equation} 
By doing this, we exclude $\sim$30\% of the initial detections. Additionally, a single cluster can result in multiple detections due to the photometric errors that can make a cluster be detected in multiple redshift slices. Thus,  when two or more clusters within two consecutive redshift slices are at transverse distance less than 1.5 $Mpc$, we just merge them into the one with highest probability. 

\section{Simulations}

We performed  simulations to test the accuracy of the algorithm and quantify the completeness and purity of the results. 

We created catalogues of galaxies, mimicking the observed properties of the simulated survey. For each richness, we simulated a 1 degree field with a cluster at z=0.1, a 1 degree field with two clusters at z=0.2 and z=0.3 and a 1 degree field with nine clusters at z=0.4 to z=1.2. We simulated nine different richnesses: $\Lambda_{CL}=$10, 20, 30, 40, 50, 75, 100, 150 and 200.

To mimic photometric redshift errors, we spread each cluster's galaxy population in a Gaussian  with  $\sigma =0.01(1+z_c)$ and $\mu =z_c$. The total number of galaxies is obtained assuming that the cluster luminosity function has a slope $\alpha$ and characteristic magnitude $M^*$ of -1.05 and -20.44 respectively.  We set the detection magnitude limit as the one measured from the simulated survey. The density distribution follows a Plummer profile as in equation (\ref{eq:plummer}). We spread the magnitudes according to different luminosity functions for different spectral types as in \cite{nakamura03} extrapolated to higher redshift. We added to each magnitude a photometric error measured from the simulated survey.

\begin{figure}
\centering
\includegraphics[clip,angle=90,width=1.0\hsize]{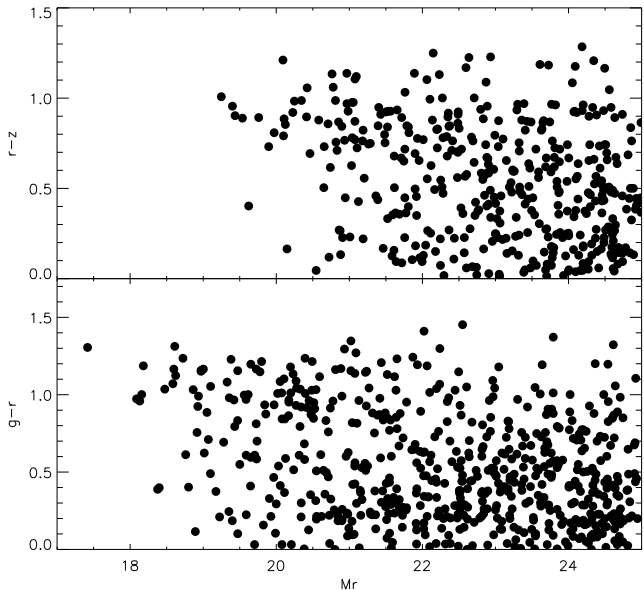} 
\caption{Upper plot: $r-z$ CMR  for a z=0.5 simulated cluster with $\Lambda_{CL}$=100. Bottom plot: $g-r$ CMR  for a z=0.2 simulated cluster with $\Lambda_{CL}$=100.}
\label{fig:CMsimu}
\end{figure}

The distribution of the two colors that sample the 4000$\AA$ break at low and high redshift were generated by using the empirical cluster color distribution by  \cite{baldry04}, including noise. We shifted their colors to our estimated colors obtained from the template spectra set of E/S0, Sbc, Scd, and Irr by \cite{coleman80} and of starbursting galaxies SB3, SB2 by \cite{kinney96} and an empirical slope from one of the well-known galaxy clusters in the survey, as explained in section 2.2. In  Figure \ref{fig:CMsimu}, we show the simulated $g-r$ and $r-z$ CMR for a  $\Lambda_{CL}=100$ cluster at redshift $0.2$ and $0.5$ respectively.

Then, we embedded these simulated clusters into a distribution of field galaxies. In order to add the most realistic background field galaxies in the simulation, we used the properties of the background field galaxies in the real survey data. To obtain the survey background distribution, we detected clusters in the survey and masked them. Then, we redistributed the positions of the remaining galaxies by  assuming an angular two-point correlation function similar to the galaxies in the real universe. We used the Rayleigh-Levy galaxy pair separation distribution:

\begin{equation} P(>\theta) = (\theta/\theta_0)^{-d} \textrm{if} \, \theta \ge \theta_0\end{equation}

We used $\theta_0=3.19''$ and $d=0.8$  extracted from the work by \cite{grazian06} for the GOODs Survey. Following the approach of \cite{postman96,postman02} and \cite{menanteau09}, we allow up to seven galaxies to be so distributed about a given center, following a random walk with this angular distribution. Once the seven galaxies are generated, we choose at random a different center until the number of galaxies matches the observed number, normalized to match the simulated survey counts. The original magnitudes, morphological types, colors and photometric redshifts of the field galaxies are preserved.  Only the positions are changed.

\subsection{Completeness and purity rates}

We examined which simulated clusters were detected and which detected clusters were not simulated. We define a recovered cluster as one which was simulated and then detected.

We define completeness as the rate of recoveries, that is, the ratio of recovered clusters to the total number of clusters simulated, and we define the purity as the ratio of recovered clusters to the total number detected.

\begin{figure}
\centering
\includegraphics[clip,angle=90,width=1.0\hsize]{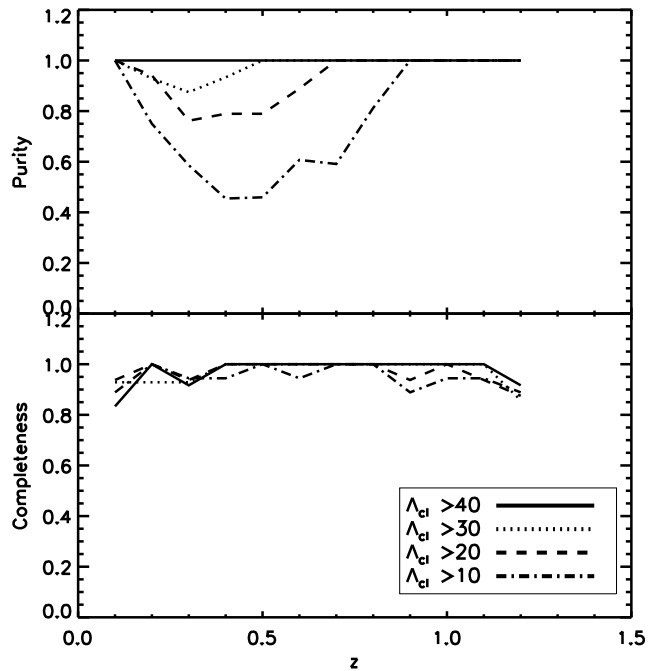} 
\caption{Completeness (bottom plot) and purity rate (top plot) for the simulated clusters with  $\Lambda_{CL}$ greater than 10, 20, 30 and  40 (solid, dotted, dashed and dotted-dashed line respectively).}
\label{fig:completeness}
\end{figure}
 
In Figure  \ref{fig:completeness}, we show the completeness and purity respectively as a function of redshift and $\Lambda_{CL}$.  The completeness is $>$90\% for all richnesses, at all redshifts for z $\le$ 1.2, the highest redshift simulated. The purity is over $\approx$ 80\% for clusters  with $\Lambda_{CL} \ge 20$ for z $\le$ 1.2. In other words, galaxy clusters richer than $ \Lambda_{CL} \ge 20$ are almost 100\% recovered, with at most 20\% of extra false positive detections for z $\le$1.2. 

Additionally, we performed different sets of simulations to prove the robustness of these simulations. We have tested possible observational effects such as contamination or misidentification of close clusters in projected and redshift space, deviations from the simulations density profile or different choices of the cluster and core radius. Simulations show that most of these effects are negligible, proving the robustness of the method.  The only non-negligible effect is if the real profile of the cluster is too different from the one employed to carry out the fit, which causes a maximum 10\% decrease in purity for poor ($\Lambda_{CL} \le 30$) clusters when the true profile is a Navarro-Frenk-White \citep{navarro97} or a random tophat profile.

In view of these results, to keep the completeness and purity rates over 80\%, we will consider real clusters detections those with $\Lambda_{CL} \ge 20$ for  z $\le$ 1.2. 

\subsection{Effect of the prior}

The universality of the red sequence in clusters has not been proved. According to the main scenarios of galaxy formation \citep{white91}, we would expect to find clusters without red sequence in the low mass regime or high redshift range. In order to take those clusters into account, we performed some sanity checks.

\begin{figure}
\centering
\includegraphics[clip,angle=90,width=1.0\hsize]{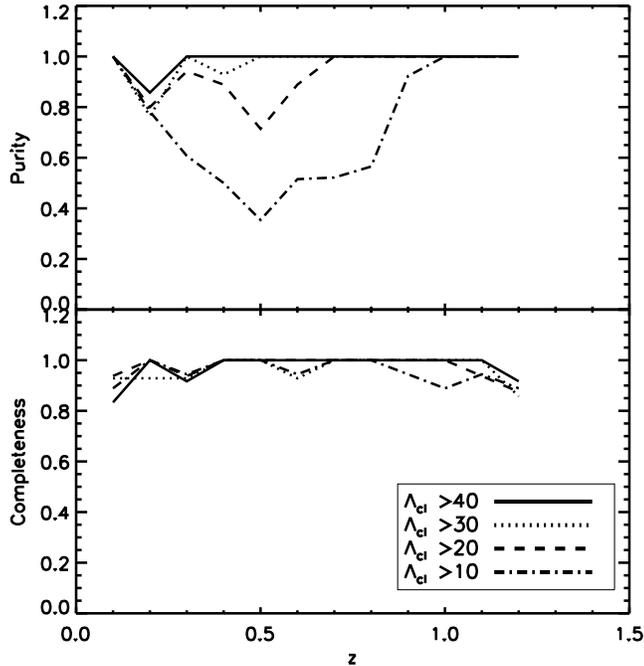} 
\caption{Completeness (bottom plot) and purity rate (top plot) for the simulated clusters with  $\Lambda_{CL}$ by using only the likelihood. The meaning of the lines are the same as in Figure \ref{fig:completeness}.}
\label{fig:completenessL}
\end{figure}

First of all, we checked how well we are recovering the galaxy clusters without the introduction of the prior. In Figure \ref{fig:completenessL}, we show the same completeness and purity fraction as in Figure \ref{fig:completeness}, but  just for the likelihood. The results are very similar. The completeness of the detections is approximately the same as when using the whole probability. We find a completeness higher than 90\% for all richnesses and  redshifts.  The purity decreases by less than $\sim$10\% for z$<$0.6 and  $\Lambda_{CL}\le$20 . Therefore, we do see that for a sample of galaxy clusters showing with a predefined red sequence \emph{for all of them}, the results of using the prior or not would be very similar.

\begin{figure}
\centering
\includegraphics[clip,angle=90,width=1.0\hsize]{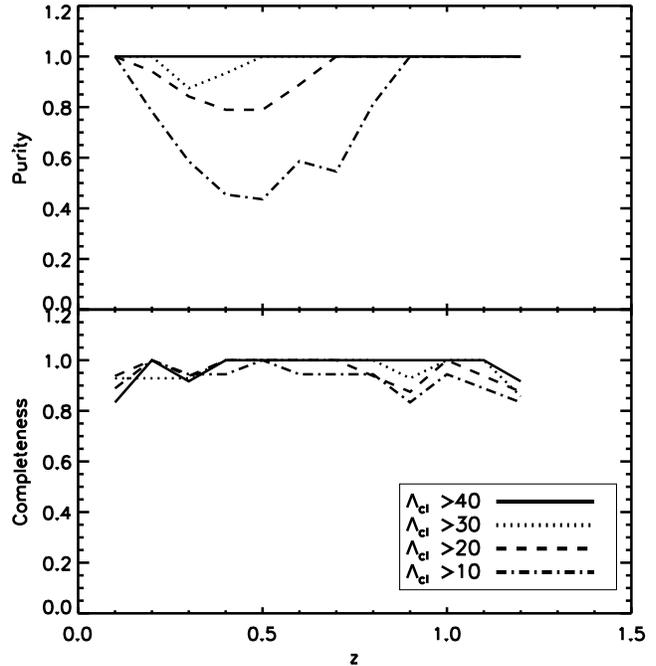} 
\caption{Completeness (bottom plot) and purity rate (top plot) for the simulated mix of clusters. The meaning of the lines are the same as in Figure \ref{fig:completeness}.}
\label{fig:completenessmix}
\end{figure}

\begin{figure}
\centering
\includegraphics[clip,angle=90,width=1.0\hsize]{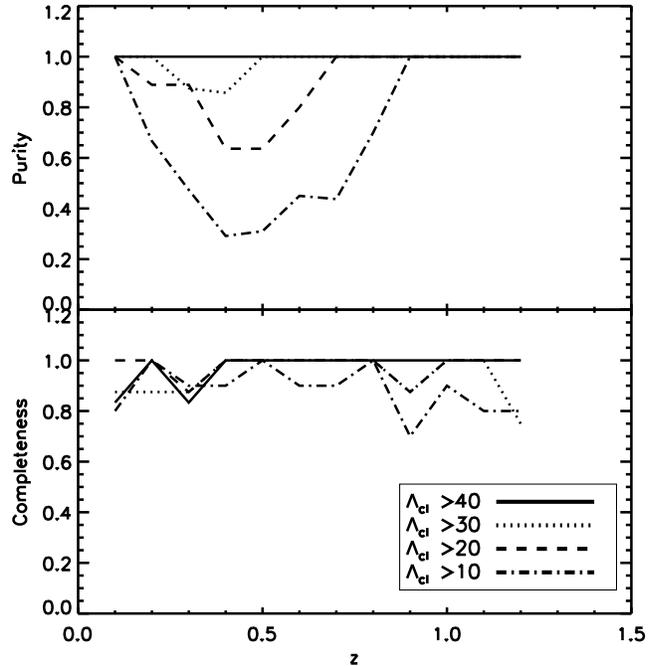} 
\caption{Completeness (bottom plot) and purity rate (top plot) for the clusters without red sequence of the simulated mix of clusters. The meaning of the lines are the same as in Figure \ref{fig:completeness}.}
\label{fig:completenessonlymix}
\end{figure}

Second, we consider a more realistic \emph{mix} of clusters, i.e: galaxy clusters with and without red sequence in the same simulation. We performed another set of simulations considering half of the clusters with and half of the clusters without red sequence. Since the background probability is calculated from all the galaxies in the survey including those in clusters, the background could decrease slightly due to the lower values of probability in the galaxy cluster without red sequence and, therefore, we might have obtain more spurious detections. In Figure  \ref{fig:completenessmix}, we show the completeness and purity rates of the whole distribution and in Figure  \ref{fig:completenessonlymix}, the same but just for the clusters \emph{without} the red sequence. The results are very similar for the whole distribution of clusters with only the completeness decreasing $<$10\% as redshift increases. However, when we consider the galaxy clusters \emph{without red sequence} the purity decreases to 60\% for galaxy clusters with $\Lambda \le 20$.  In this case, we define purity as the ratio of recovered clusters without a red sequence to the total number detected excluding the total number detected with a red sequence.

\begin{figure}
\centering
\includegraphics[clip,angle=90,width=1.0\hsize]{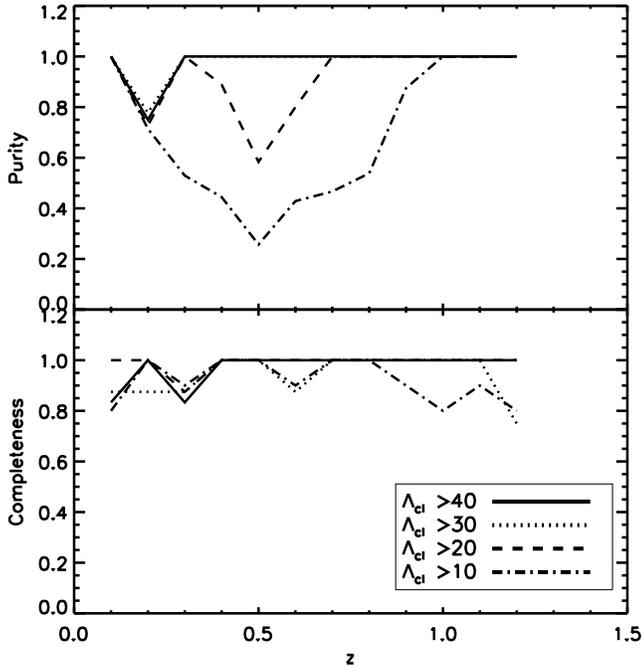} 
\caption{Completeness (bottom plot) and purity rate (top plot) for the clusters without red sequence of the simulated mix of clusters, using only the likelihood. The meaning of the lines are the same as in Figure \ref{fig:completeness}.}
\label{fig:completenessLonlymix}
\end{figure}

We show in Figure \ref{fig:completenessLonlymix}, the results of using just the likelihood for those clusters without a red sequence generated in the mixed simulation. The rates keep approximately the same since those galaxy clusters have been simulated without a red sequence by construction. In summary, the inclusion of the prior in the main probability and the consideration of the photometric redshifts allow the method to obtain high purity and completeness rates, being able to detect almost every cluster richer than $\Lambda_{CL}\le20$ for z $\le$ 1.2 with or without red sequence with a very low contamination rate (20\% at most).

\subsection{Cluster property recovery}

\begin{figure}
\centering
\includegraphics[clip,angle=90,width=1.\hsize]{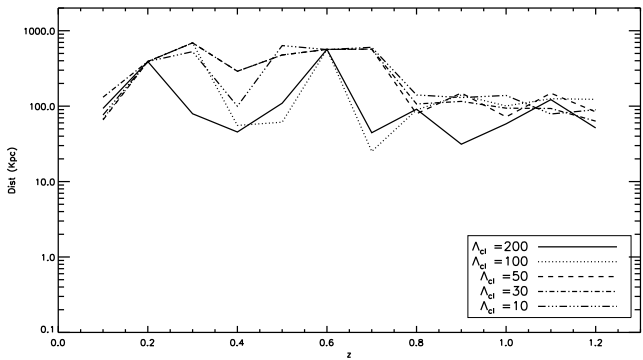} 
\caption{Position offset in the recovery of the clusters as a function of the redshift and $\Lambda_{CL}$.}
\label{fig:offpositions}
\end{figure}

\begin{figure}
\centering
\includegraphics[clip,angle=90,width=1.\hsize]{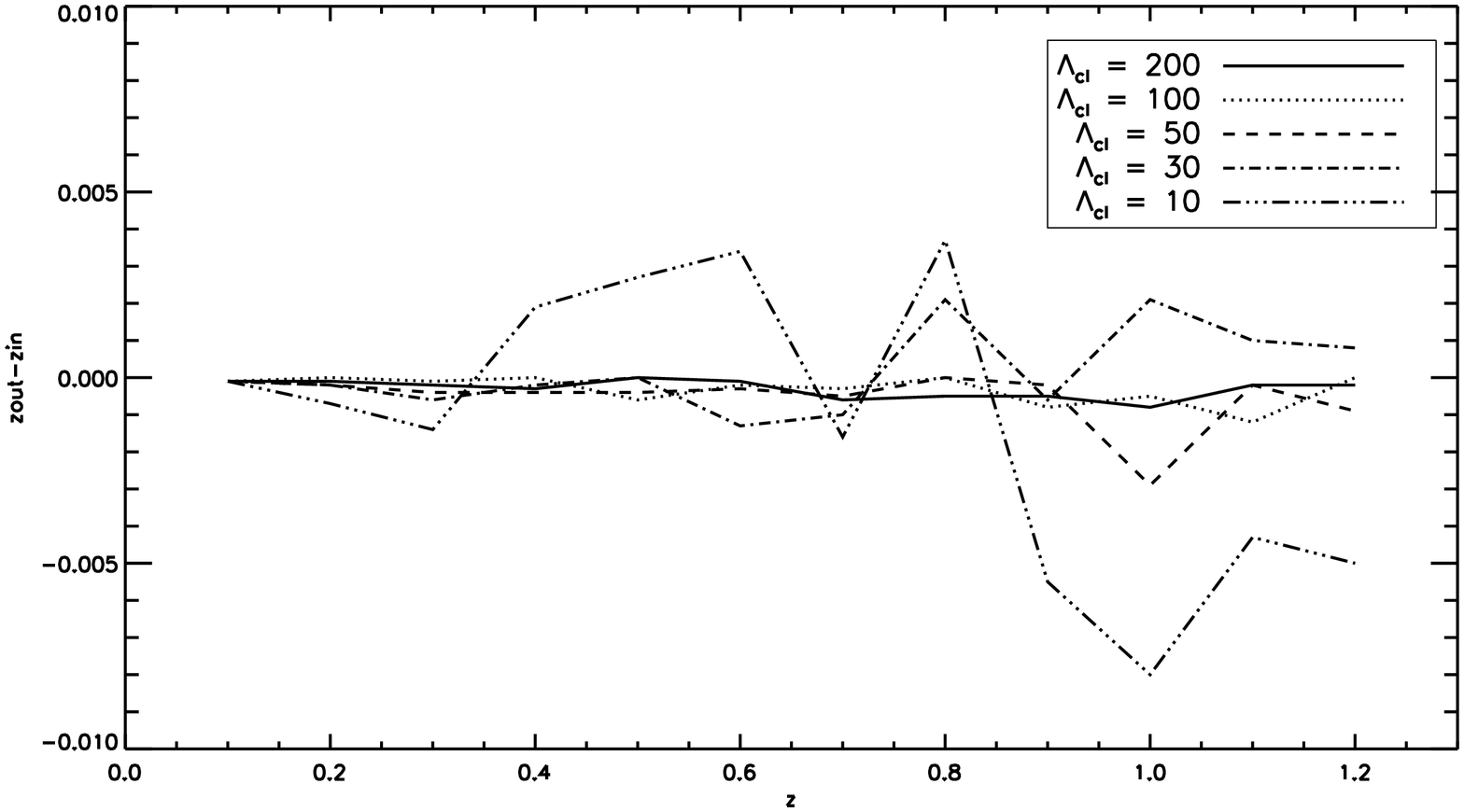} 
\caption{Redshift offset in the recovery of the clusters as a function of the redshift and $\Lambda_{CL}$.}
\label{fig:offredshift}
\end{figure}

\begin{figure}
\centering
\includegraphics[clip,angle=90,width=1.\hsize]{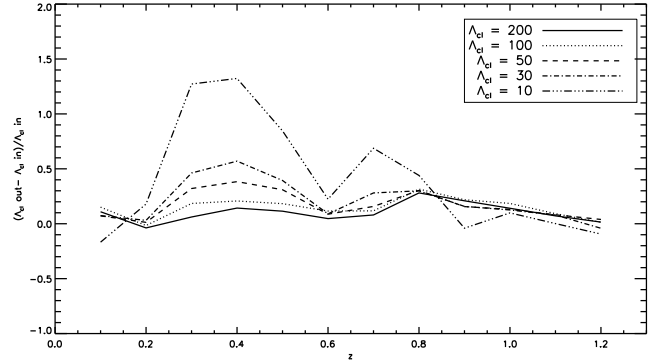} 
\caption{$\Lambda_{CL}$ richness offset in the recovery of the clusters as a function of the redshift and $\Lambda_{CL}$.}
\label{fig:offrichness}
\end{figure}

We compare the simulated cluster properties with the recovered ones. In Figures  \ref{fig:offpositions}, \ref{fig:offredshift} and \ref{fig:offrichness}, we show the offset in the recovery of the position, redshift and richness respectively, as a function of redshift and richness. 

The positions are quite well recovered, with offsets from 100 kpc for the most massive clusters up to 500 kpc for the least massive ones. Redshifts are very well determined (between 0.001 difference for the most massive clusters and up to 0.008 for the least massive clusters). Finally, the richnesses are slightly overestimated, especially for very poor clusters. This behavior is expected as the output cluster is recovering some contamination from the field galaxies.  

In general, we find a good agreement between the properties of the simulated and recovered clusters, only finding a higher misalignment and richness overestimation for the very poor clusters ($\Lambda_{CL} \le$ 20).

\section{Clusters in CFHTLS-Archive Research Survey}

\subsection{Description of the sample}

The Canadian-France-Hawaii Telescope Legacy Survey (CFHTLS) Deep and Wide consists of 4 and 171 degree squares respectively in five optical bands ($ugriz$). The completeness of this survey is up to 26 (CFHTLS-D) and 24.5 (CFHTLS-W) in r band.  The combination of both width and depth of this survey makes it  an excellent sample to test our algorithm to detect clusters. 

\cite{erben09} introduced the CFHTLS- Archive-Research Survey (CARS), which is based on the public archive images from the Wide CFHTLS. They created multi-color catalogues, and estimated photometric redshifts with BPZ \citep{benitez00} in almost 37 square degrees, 21, 5 and 11 square degrees in W1, W3 and W4 respectively. 

The BPZ provides a quality indicator called odds, which is defined as the integral of the redshift probability distribution within a given interval centered on the recovered Bayesian photometric redshift. This parameter is useful to discard objects with unreliable photo-z. We selected all galaxies with odds parameter $> 0.70$  as a compromise between keeping the maximum number of galaxies in the survey and avoiding catastrophic outliers. We consider as galaxies those objects with stellar galaxy parameter less than 0.2 and magnitude less than 21 and all of them for magnitude greater than 21.

\begin{figure}
\centering
\includegraphics[clip,angle=90,width=1.0\hsize]{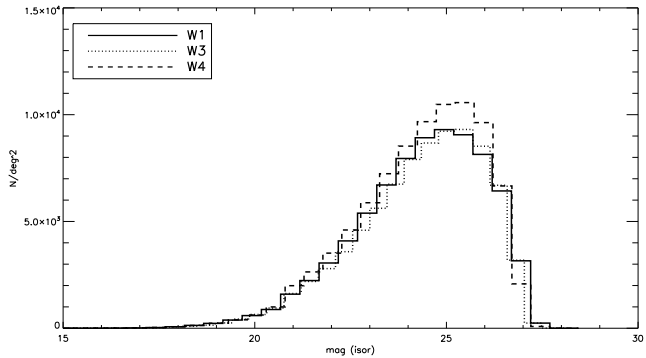} 
\caption{CARS Magnitude distribution}
\label{fig:magCARS}
\end{figure}

\begin{figure}
\centering
\includegraphics[clip,angle=90,width=1.0\hsize]{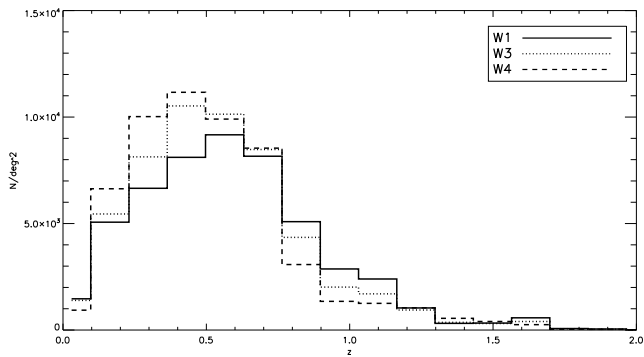} 
\caption{CARS Photometric redshift distribution}
\label{fig:zCARS}
\end{figure}

In Figures  \ref{fig:magCARS} and \ref{fig:zCARS}, we show the galaxy magnitude and photometric redshift distribution for W1, W3 and W4. The redshift distribution peaks at $\sim$  0.6 and the magnitude distribution at  $m_r \sim$ 25, this being  the limit we are going to consider as the completeness of the data.

\subsection{Cluster detection}

We detected galaxy clusters in W1, W3 and W4 from the CARS catalogues. We obtained 735, 171 and 340 clusters with  $\Lambda_{CL} \ge 20$ and z $\le$ 1.2. This corresponds to 35, 34.2 and 30.91 per square degree. We now compare those cluster candidates with different detections found by other works. We should keep in mind though that the way that each work merges the detections is different in each case. As an example, if we assumed 1 Mpc instead of 1.5 Mpc as the maximum distance for the detections to be considered the same, we would end up with 42.38, 40.4 and 38 detections per square degree. Or, if we did not apply any merging procedure to the detection, we would obtain 51.24, 52.6 and 46.55 detections per square degree. This is an issue when using galaxy cluster counts for cosmological purposes.

\subsubsection{Comparison with Olsen et al. 2007}

\cite{olsen07} (hereafter O07) detected 46 clusters in Deep1, which overlaps with one square degree in W1. We detected 53 clusters in the common area of W1. We matched our detections to their detections by using a Friend of Friends (FoF) Algorithm \citep{huchra82,botzler04}. For each of our detections, we searched for their O07 friends. These friends have to fulfill that they are within a comoving distance of 4 Mpc considering the errors in the redshift measurement. We calculated the redshift errors by measuring the width of the Gaussian that we fit to the photometric redshift distribution in order to obtain $z_{est}$. Then, we searched for friends of friends until no more friends are found. Of the final list of friends, we select the one with the closest photometric redshift to the original detection. Finally, we consider a cluster to have a counterpart in B if their \emph{best friend} is within a comoving distance of 4 Mpc. By \emph{matching A with B} we are asking whether each cluster in A has at least one counterpart in B, so it is not the same as matching B with A. We allow a cluster from A to match more than one cluster in B since every different study follows different merging algorithms.

\begin{figure}
\centering
\includegraphics[clip,angle=90,scale=0.5]{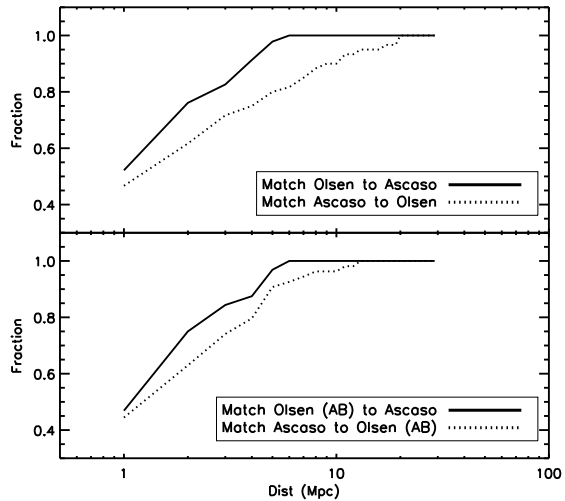} 
\caption{Comoving distance between detections versus fraction of detections that matches O07 to this work (solid line) and this work to O07 (dotted line). The upper plot considers the whole O07 sample, whereas the bottom plot only considers the A and B systems by O07.}
\label{fig:matchesOlsen}
\end{figure}

In Figure \ref{fig:matchesOlsen}, we show the relation between the comoving distance and fraction of matched detections. In the upper plot, the whole O07 detections are included whereas in the bottom plot, only the O07 subsample of 32 clusters, which excludes all the C or D systems. (In O07's notation, C systems are those which do not show any clear galaxy overdensity at examining the optical image, D grade is assigned to those systems which were detected due to artefacts such as an image edge or a lack of masking. Additionally, A systems refer to clear concentration of galaxies with similar colors whereas B systems are defined to appear as an overdensity of galaxies, less concentrated than A systems or without an obvious color concentration). We can see that with a 4 Mpc distance we find that 85\% of our sample matches O07, whereas only 70\% of O07 matches our sample. However, if we just consider the subsample of A and B systems by O07, we find a fraction of 80\% matching within 4 Mpc  for both samples.

\begin{figure}
\centering
\includegraphics[clip,angle=90,scale=0.4]{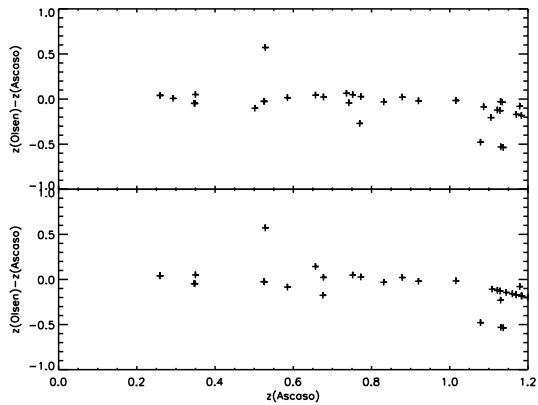} 
\caption{Redshift difference between a cluster candidate found in this work and their matched counterpart from  O07, with respect to the redshift estimated in this work. The top panel refers to the matches using all the O07 catalogue, whereas the bottom panel excludes the C and D systems from O07 catalogue}
\label{fig:zolsen}
\end{figure}

In Figure \ref{fig:zolsen}, we plot the difference between the redshift  of the matched clusters with respect to the redshift estimated in this work.  The top panel refers to the matches using all the detections in O07, whereas the bottom panel only refers to the detections excluding C and D systems in O07.  We find a good agreement in redshift at any redshift with a slight tendency of $z_{Olsen}-z_{Ascaso} <$0 at $z_{Ascaso} > 1$. 

\begin{figure}
\centering
\includegraphics[clip,angle=90,scale=0.4]{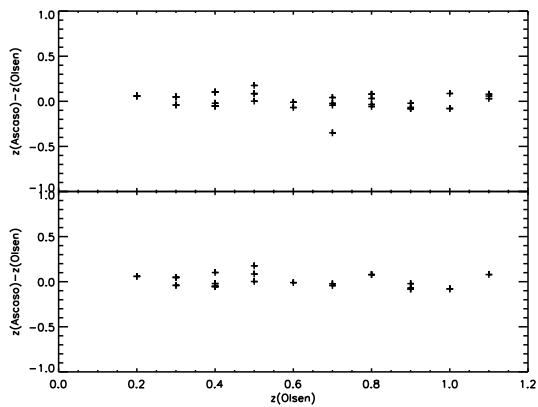} 
\caption{Redshift difference between a cluster candidate from O07 and their matched counterpart in this work, with respect to the O07 redshift. The top panel refers to the matches using all the O07 catalogue, whereas the bottom panel excludes the C and D systems from  O07 catalogue. Note the reversal of sign of the residuals compared to Fig. \ref{fig:zolsen}.}
\label{fig:zolsen2}
\end{figure}

Additionally, we matched the catalogs in the reverse order. By doing this, we want to know how pure our detections are with respect to the other catalogue. We used the same criteria as previously explained to create the matches. We show the difference between the redshift of the matched clusters with respect to the O07 redshift estimations in Figure \ref{fig:zolsen2}. As before, the top panel refer to the matches using all the detection in O07, whereas the bottom panel only refers to the detections excluding C and D systems in O07. We see a very good agreement between both redshifts at any redshift. In the rest of the paper, we include the C and D systems.

\begin{figure}
\centering
\includegraphics[clip,angle=90,scale=0.5]{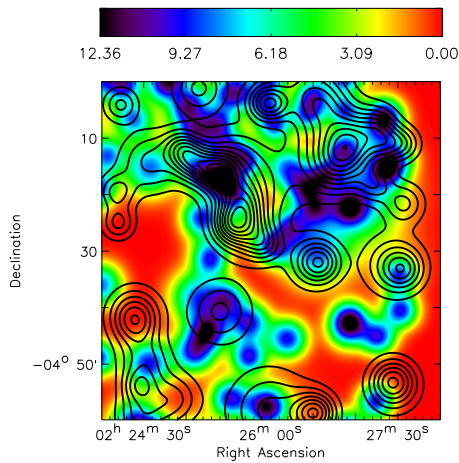} 
\caption{Density maps of the cluster detections in this work in the area of CARS that overlaps with the Deep 1 field (color maps). The black contours correspond to the level curves of the density maps for the O07 catalogue.}
\label{fig:W1p2p3}
\end{figure}

In order to take into account the possible biases in the way that each work merges their detection by using different criteria, we  created density maps of the cluster detections. Each detection is represented by a a Gaussian centered on the position of the cluster candidate, with a width of 1 Mpc radius at the redshift of the cluster. In Figure  \ref{fig:W1p2p3}, we show the density map for the detections found in this work with a color map. Additionally, we overplot the level curves of the  density maps for the cluster candidates by  O07  in Deep 1. We see a very good agreement between both detections. The main two prominent structures in the upper left and upper right are well identified in both cases. We also identify the four subclumps at the bottom. The only visible difference is the more elongated structure that O07  finds in the bottom left of the image and that we do not detect.

\begin{figure}
\centering
\includegraphics[clip,angle=90,scale=0.5]{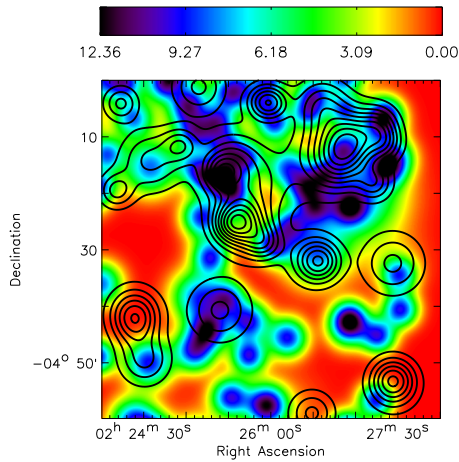} 
\caption{The same as in Figure \ref{fig:W1p2p3} but the black contours correspond to  the level curves of the density maps for the  O07 systems classified as A or B.}
\label{fig:W1p2p3AB}
\end{figure}

However, when we overplot the level curves of the density maps only for the A and B systems, or those which have been visually classified as having a clear  cluster-like structure, this elongation becomes smaller as shown in Figure \ref{fig:W1p2p3AB}. We find very similar  structures as found in O07, even if the latter are based on Deep 1 data, which is almost $\sim$ 1 magnitude fainter than the Wide Fields used in this work.

\subsubsection{Comparison with  Adami et al. 2010}

Recently, \cite{adami10} (hereafter A10) published detections in 19 out of 21 degrees of W1, 4 out of 5 degrees in W3 and 2 out of 11 degrees in W4. Their method consisted of an adaptive kernel technique combined with SExtractor for the detection of the structures. They used the T0004 catalogues from the CFHTLS. They also obtained photometric redshifts by using the Le Phare software \citep{ilbert06}. Thus, the direct comparison between both methods could be affected by the systematics of the different catalogues and methods used in the data.

We obtain 682, 138 and 70 clusters in the common area of W1, W3 and W4. They detected  clusters overÄ 2 and 3$\sigma$, where the $\sigma$ is estimated from SExtractor. Over 2$\sigma$, they obtain 755, 175 and 99 detections. A10's 3$\sigma$ catalogs contain 441, 130 and 31 cluster candidates. As we explained at the beginning of the section, the numbers are not directly comparable since each work uses a different merging criteria. However, our threshold appears to be somewhat closer to their 2$\sigma$ threshold than their 3$\sigma$ threshold.

\begin{figure}
\centering
\includegraphics[clip,angle=90,scale=0.5]{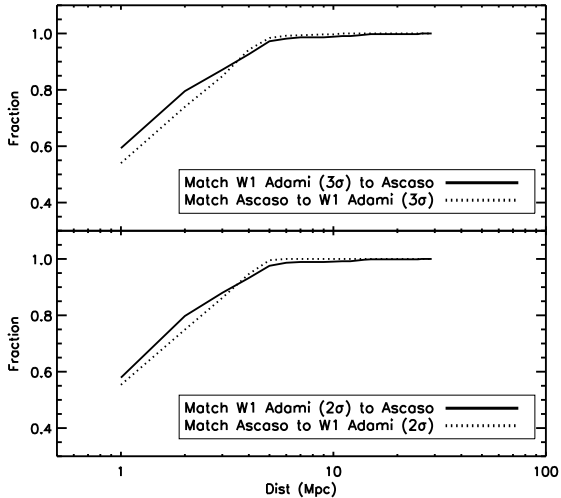} 
\caption{Comoving distance between detections versus fraction of detections that matches A10 in W1 to this work (solid line) and this work to A10 in W1 (dotted line). The upper plot considers the A10 sample in W1 over 3$\sigma$, whereas the bottom plot  considers the A10 sample in W1 over 2$\sigma$.}
\label{fig:matchesAdamiW1}
\end{figure}

\begin{figure}
\centering
\includegraphics[clip,angle=90,scale=0.5]{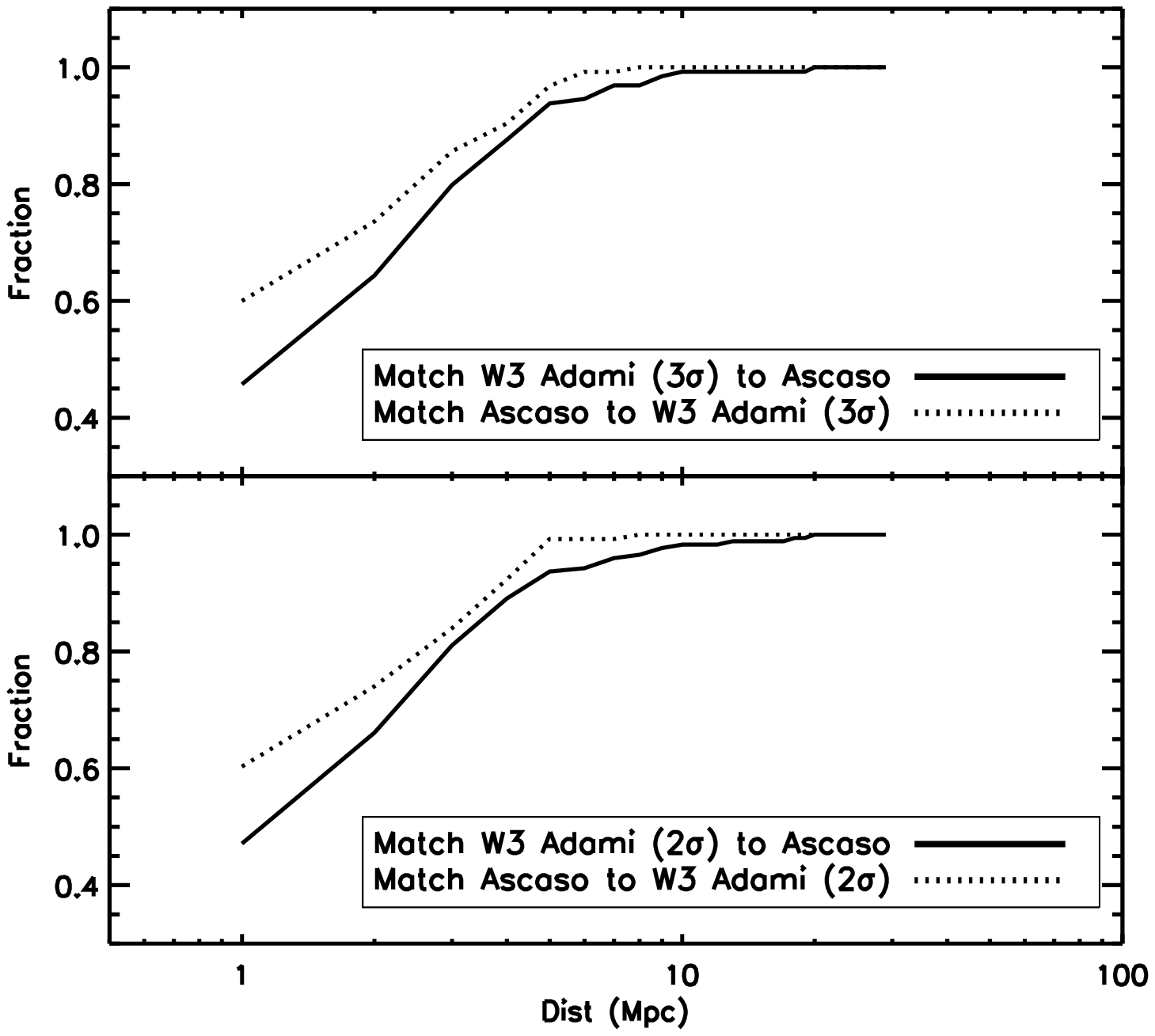} 
\caption{Comoving distance between detections versus fraction of detections that matches A10 in W3 to this work (solid line) and this work to A10 in W3 (dotted line). The upper plot considers the A10 sample in W3 over 3$\sigma$, whereas the bottom plot  considers the A10 sample in W3 over 2$\sigma$.}
\label{fig:matchesAdamiW3}
\end{figure}

\begin{figure}
\centering
\includegraphics[clip,angle=90,scale=0.5]{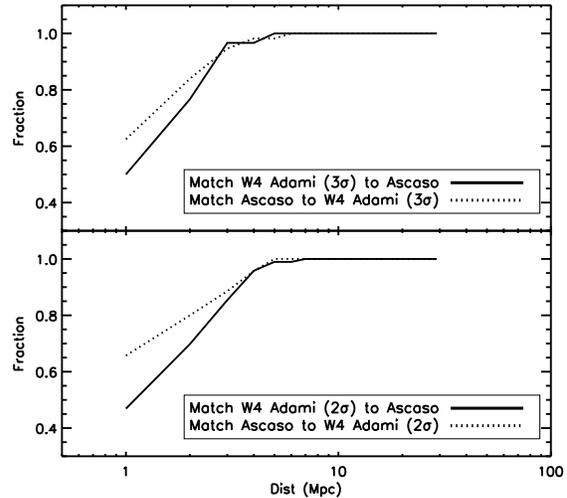} 
\caption{Comoving distance between detections versus fraction of detections that matches A10 in W4 to this work (solid line) and this work to A10 in W4 (dotted line). The upper plot considers the A10 sample in W4 over 3$\sigma$, whereas the bottom plot  considers the A10 sample in W4 over 2$\sigma$.}
\label{fig:matchesAdamiW4}
\end{figure}

We matched the detections found in this work to their detections using the same criteria as in the previous section. In Figures \ref{fig:matchesAdamiW1}, \ref{fig:matchesAdamiW3} and \ref{fig:matchesAdamiW4}, we show the maximum comoving distance for matching two detection versus the percentage of detections for the A10 sample over 3 and 2$\sigma$ respectively for the overlapping area with W1, W3 and W4. In all cases the fraction of matched detections become higher than 80\% for distances less than 4 Mpc.

\begin{figure}
\centering
\includegraphics[clip,angle=90,scale=0.4]{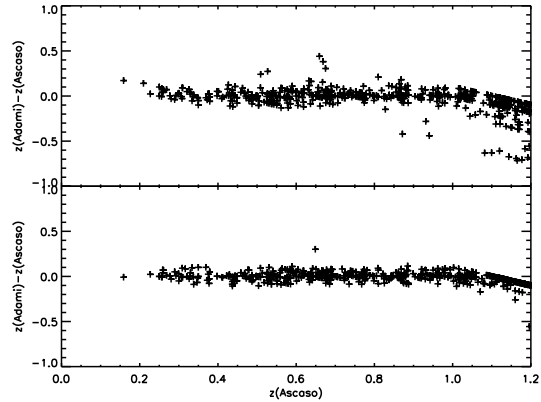} 
\caption{Redshift difference between the cluster candidates in this work and matched clusters from  A10's catalogs for the common area in W1. The upper and bottom plot are matched to the A10 detections over 3 and 2 $\sigma$ respectively.}
\label{fig:zadamiW1}
\end{figure}

\begin{figure}
\centering
\includegraphics[clip,angle=90,scale=0.4]{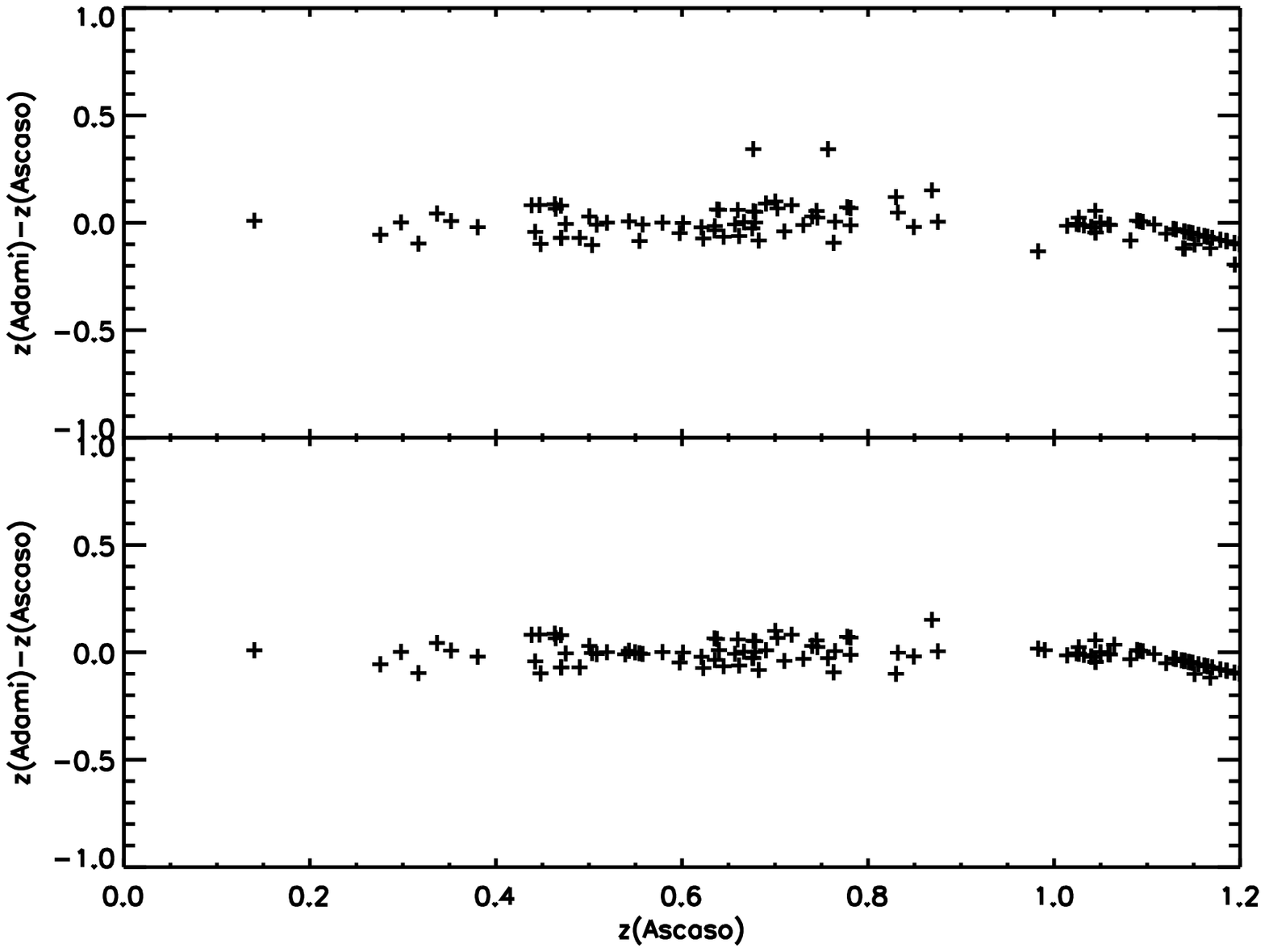} 
\caption{Redshift difference between the cluster candidates in this work and matched clusters from  A10's catalogs for the common area in W3. The upper and bottom plot are matched to the A10 detections over 3 and 2 $\sigma$ respectively.}
\label{fig:zadamiW3}
\end{figure}

\begin{figure}
\centering
\includegraphics[clip,angle=90,scale=0.4]{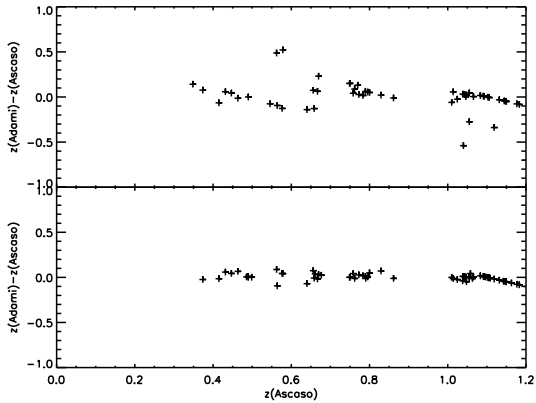} 
\caption{Redshift difference between the cluster candidates in this work and matched clusters from  A10's catalogs for the common area in W4. The upper and bottom plot are matched to the A10 detections over 3 and 2 $\sigma$ respectively.}
\label{fig:zadamiW4}
\end{figure}

In Figures \ref{fig:zadamiW1}, \ref{fig:zadamiW3} and \ref{fig:zadamiW4} we show the redshift difference between the cluster candidates in this work and the matched A10 candidates  in W1, W3 and W4 respectively over 3 and 2$\sigma$ respectively, as a function of our redshift estimate. The redshift difference is consistent with zero for redshift z$<$1.1 and $z_{Adami}-z_{Ascaso}$ tends to be systematically negative at very high $z_{Ascaso}$ ($z_{Ascaso}>$1.1). However, the dispersion is much smaller when compared to the A10 2$\sigma$ detections, in particular in W4.

\begin{figure}
\centering
\includegraphics[clip,angle=90,scale=0.4]{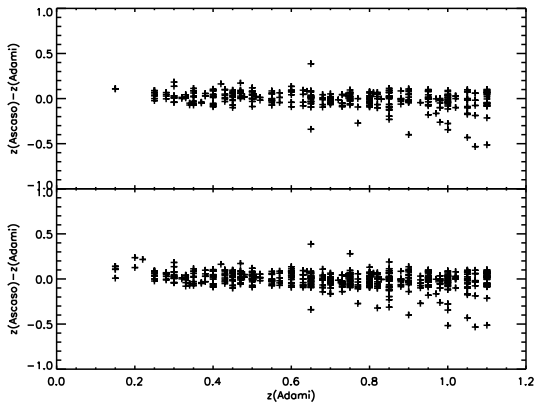} 
\caption{Redshift difference between the cluster candidates in A10 and their matched cluster candidate from this work in the common area in W1. The upper and bottom plot are detections matched to A10 catalogues over 3 and 2 $\sigma$ respectively.}
\label{fig:zadami2W1}
\end{figure}

\begin{figure}
\centering
\includegraphics[clip,angle=90,scale=0.4]{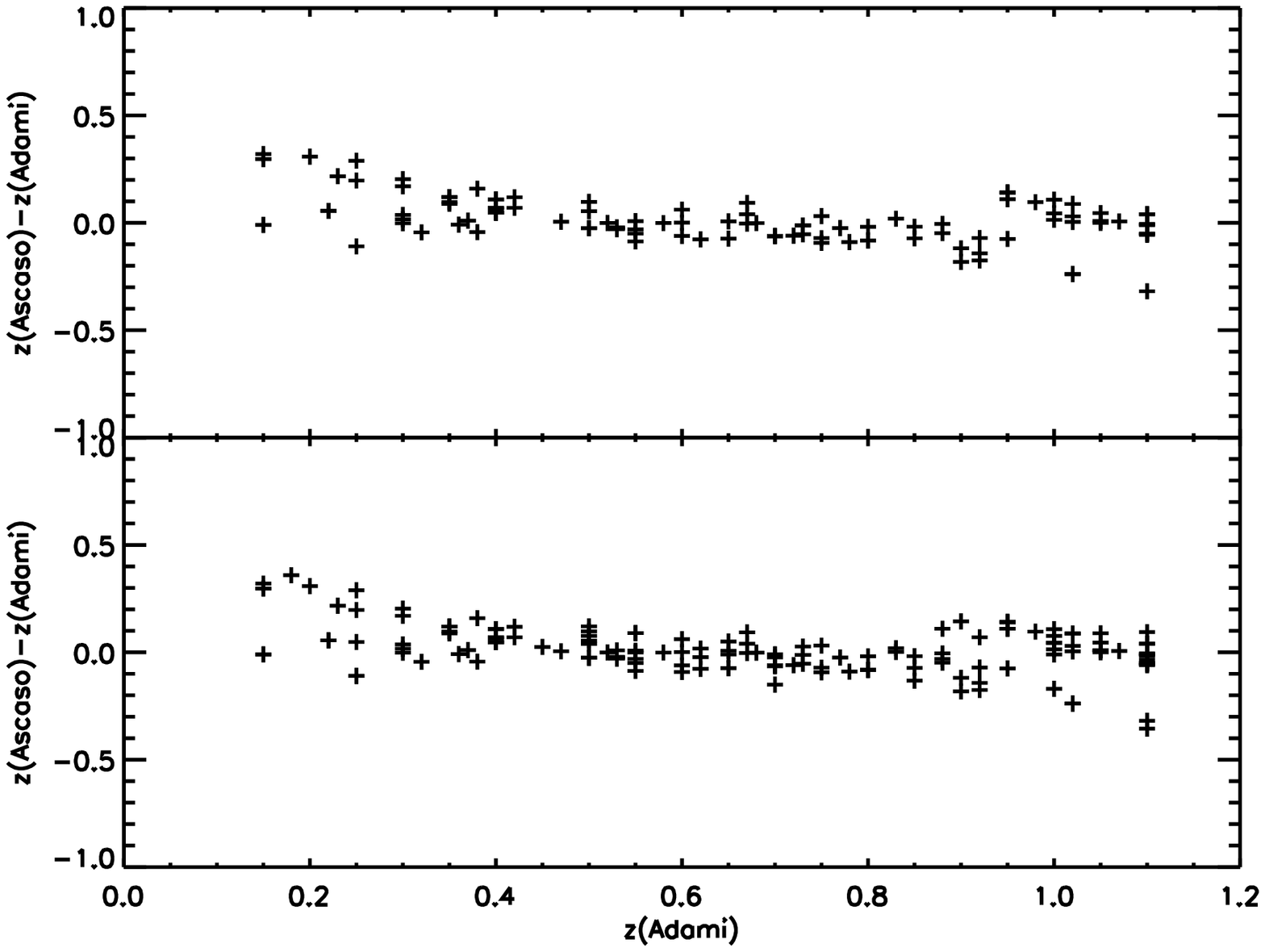} 
\caption{Redshift difference between the cluster candidates in A10 and their matched cluster candidate from this work in the common area in W3. The upper and bottom plot are detections matched to A10 catalogues over 3 and 2 $\sigma$ respectively.}
\label{fig:zadami2W3}
\end{figure}

\begin{figure}
\centering
\includegraphics[clip,angle=90,scale=0.4]{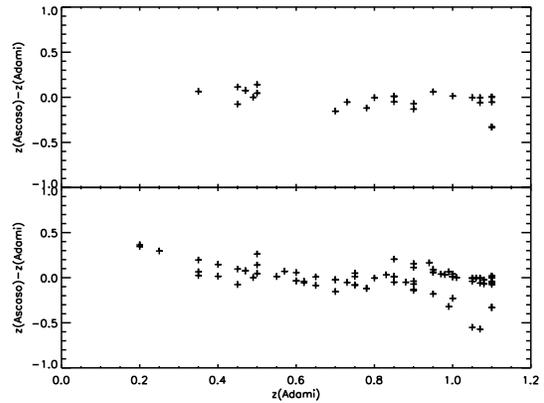} 
\caption{Redshift difference between the cluster candidates in A10 and their matched cluster candidate from this work in the common area in W4. The upper and bottom plot are detections matched to A10 catalogues over 3 and 2 $\sigma$ respectively.}
\label{fig:zadami2W4}
\end{figure}

Again, we reversed the order of matching to check on any artifacts of the matching process. In Figures \ref{fig:zadami2W1}, \ref{fig:zadami2W3} and \ref{fig:zadami2W4}, we show the redshift difference between the cluster candidates in this work that match those A10 candidates for W1, W3 and W4. 

We find some cases in all the fields where the redshift difference for the matched clusters to the A10 catalogue is negative at higher $z_{A10}>$1. There are a few cases where the difference is positive at $z_{A10}<$0.3 as in W3. Since the redshift errors are not available in A10, we are not taking them into account when matching the detection in this work to their detection. This might produce a misidentification of the right candidate. Other than that, we find a good agreement between both detection.

\begin{figure}
\centering
\includegraphics[clip,angle=90,scale=0.5]{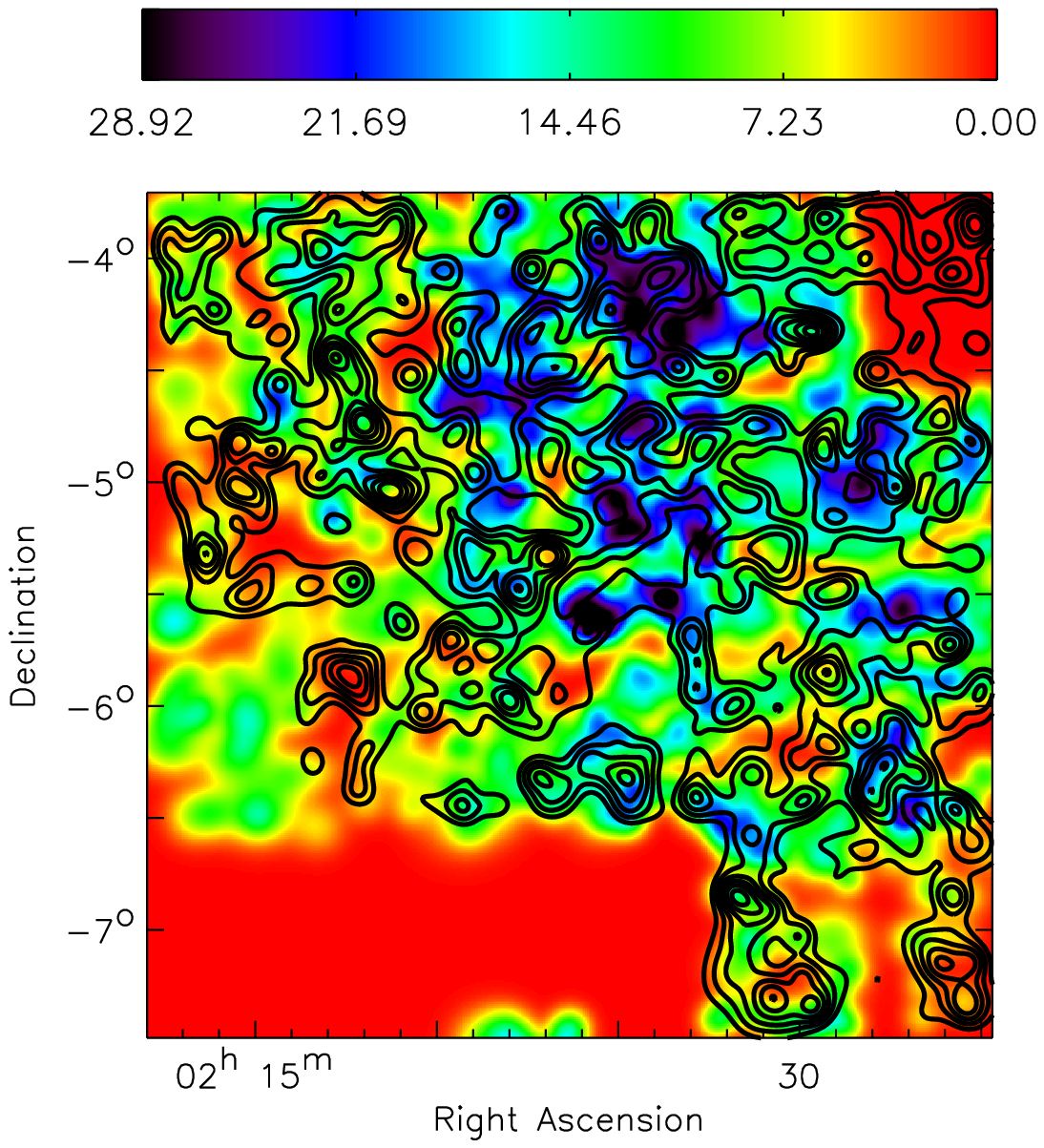} 
\includegraphics[clip,angle=90,scale=0.5]{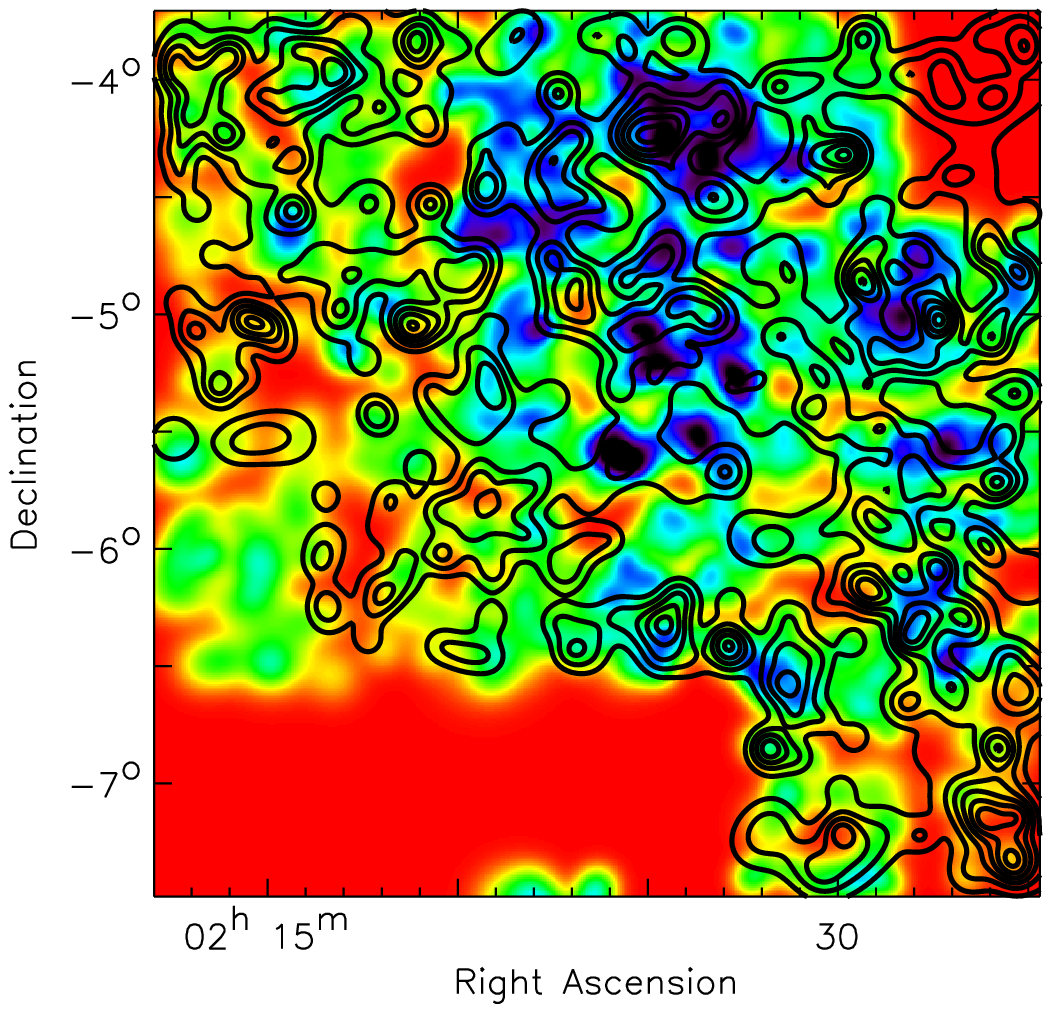} 
\caption{Density maps of the cluster detections in this work in the area of CARS that overlaps with the area analyzed by A10 in  field W1 (color maps). The black contours correspond to the level curves of the density maps from the A10 catalogue. Upper and bottom plots show the level curves over A10's 2 and 3$\sigma$ catalogs respectively.}
\label{fig:AdamiW1}
\end{figure}

\begin{figure}
\centering
\includegraphics[clip,angle=90,scale=0.5]{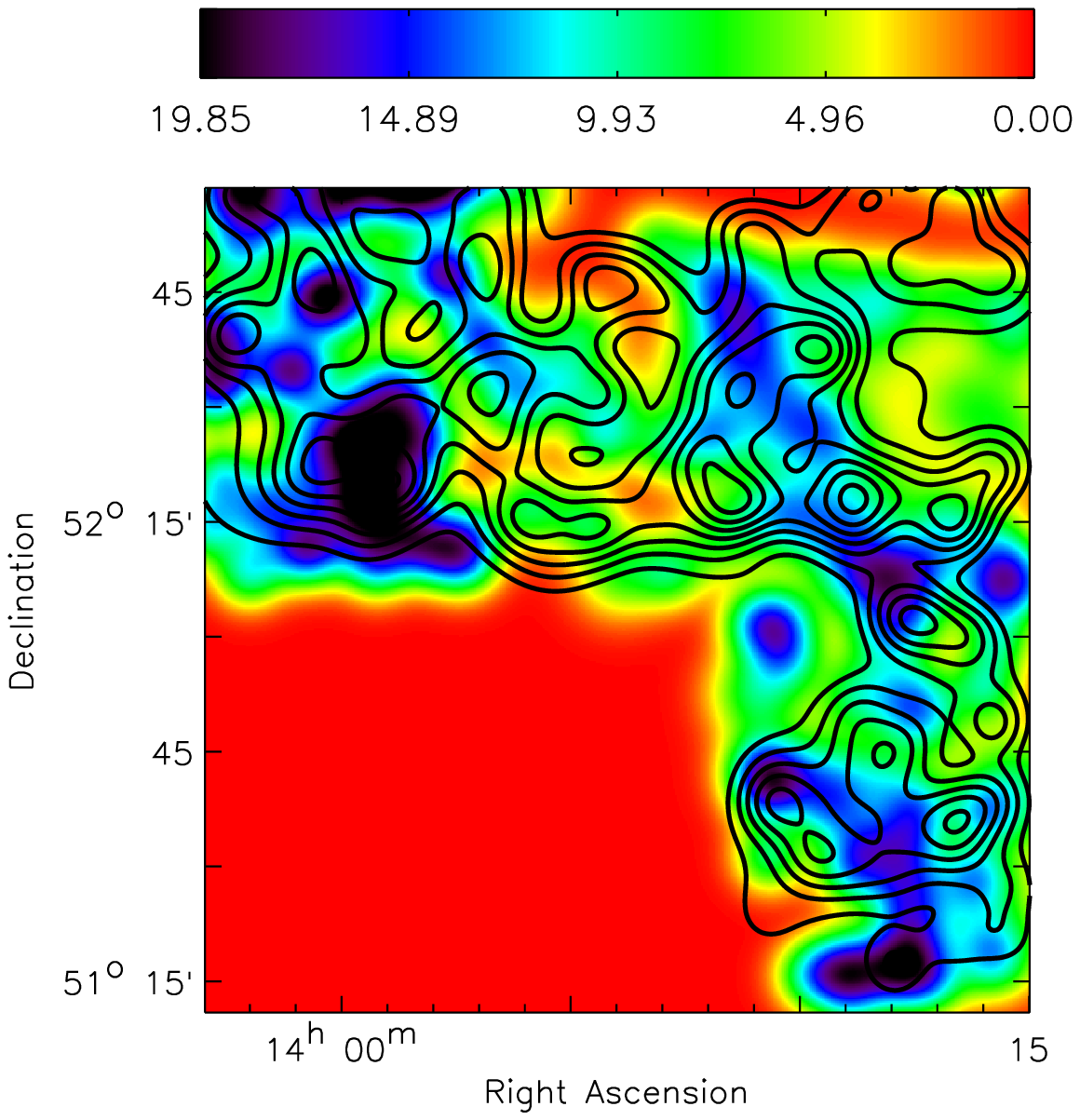} 
\includegraphics[clip,angle=90,scale=0.5]{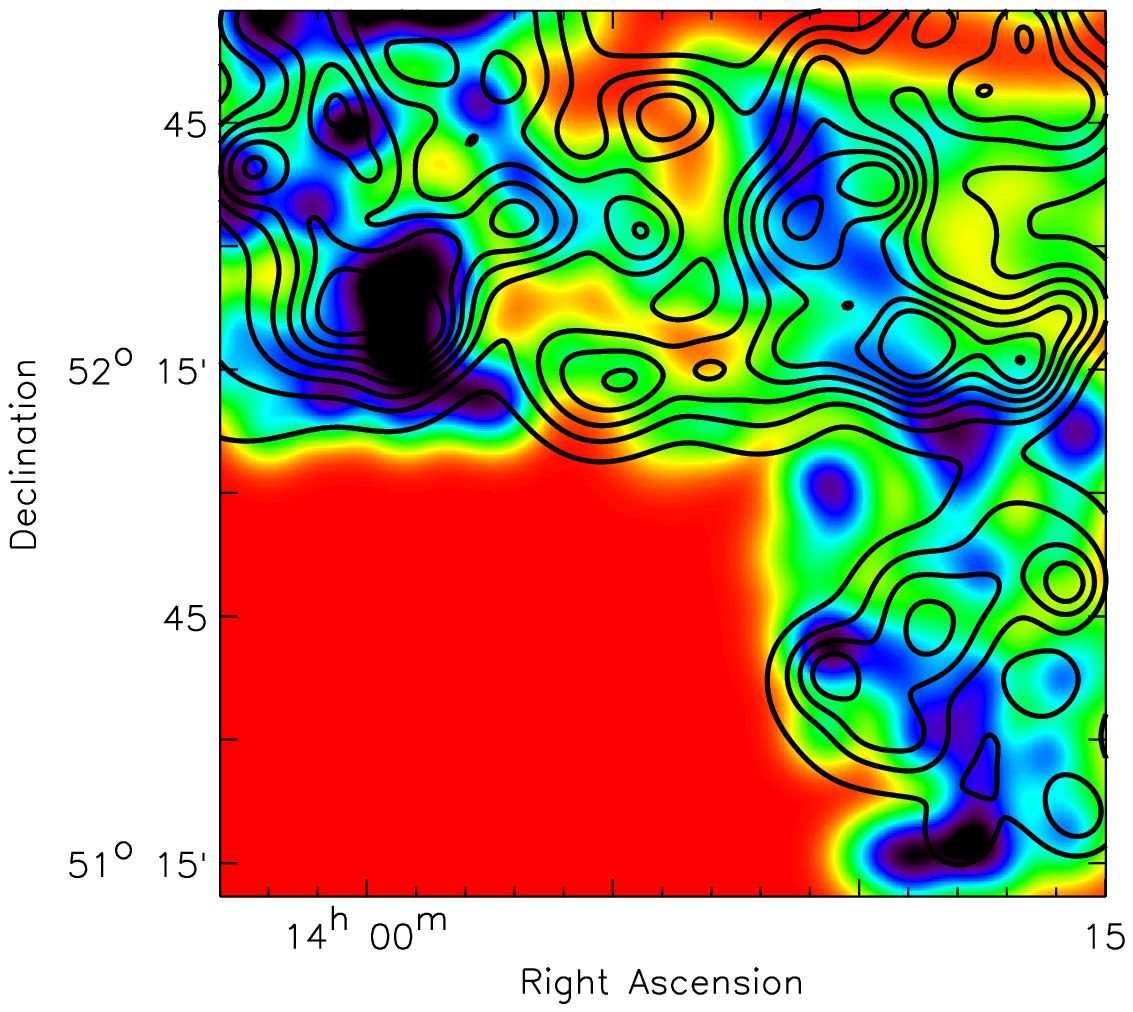} 
\caption{Density maps of the cluster detections in this work in the area of CARS that overlaps with the area analyzed by A10 in  field W3 (color maps). The black contours correspond to the level curves of the density maps from the A10 catalogue. Upper and bottom plots show the level curves over A10's 2 and 3$\sigma$ catalogs respectively.}
\label{fig:AdamiW3}
\end{figure}

\begin{figure}
\centering
\includegraphics[clip,angle=90,scale=0.5]{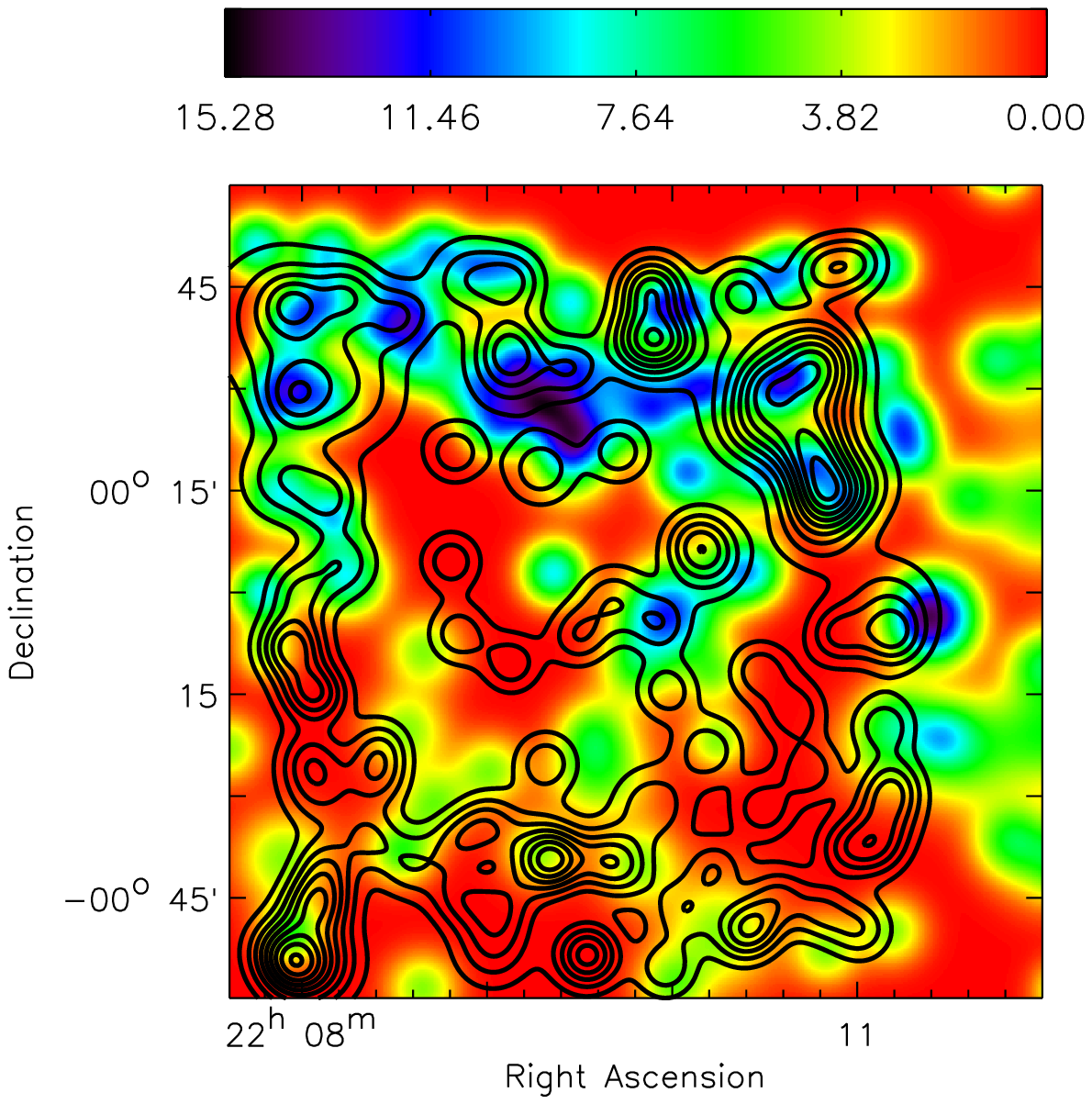} 
\includegraphics[clip,angle=90,scale=0.5]{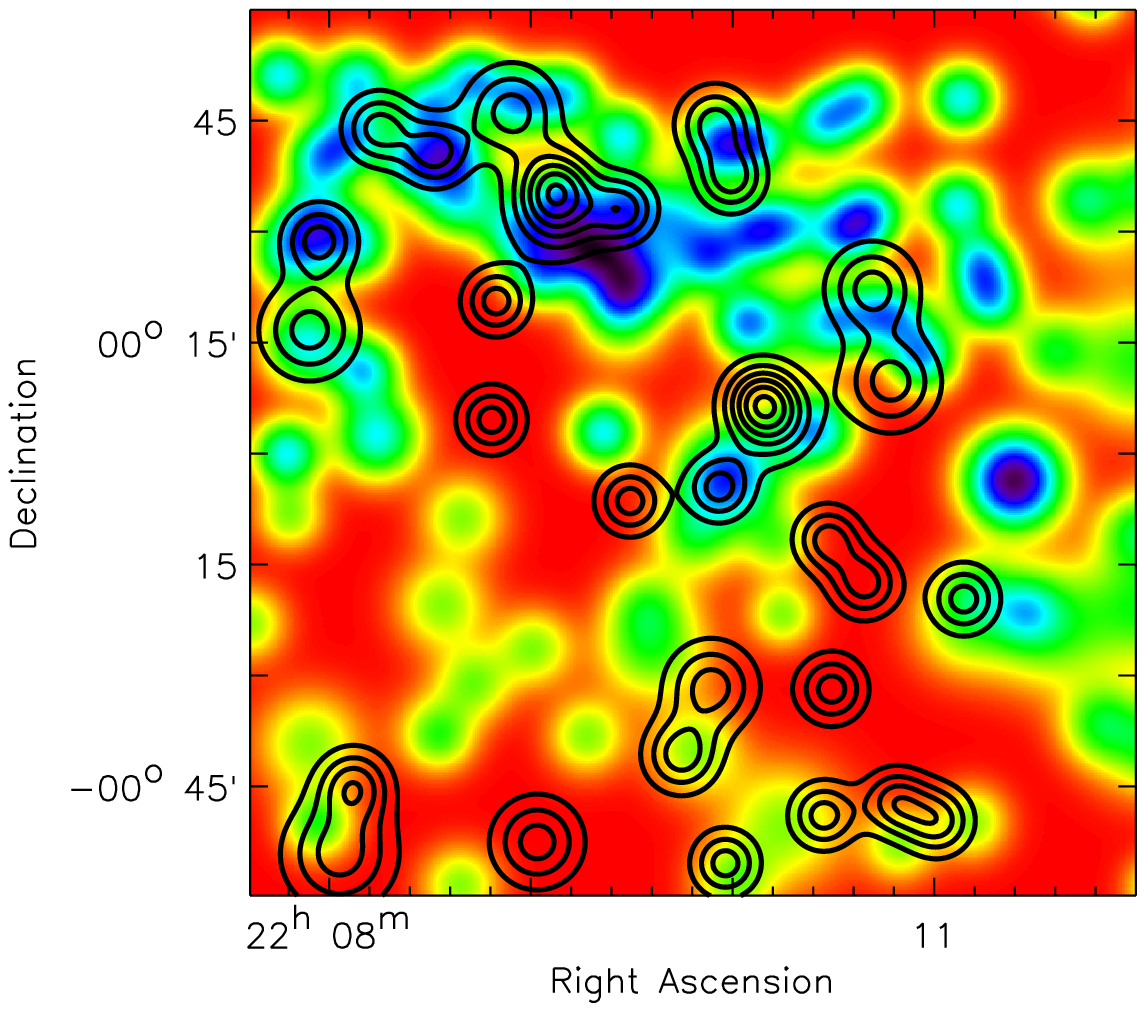} 
\caption{Density maps of the cluster detections in this work in the area of CARS that overlaps with the area analyzed by A10 in  field W4 (color maps). The black contours correspond to the level curves of the density maps from the A10 catalogue. Upper and bottom plots show the level curves over A10's 2 and 3$\sigma$ catalogs respectively.}
\label{fig:AdamiW4}
\end{figure}

As before, we created the density maps of the detections found in this work for the common area and compared them with contour levels for the detections found by A10 over 3 and 2$\sigma$ in order to avoid systematics due to different merging procedures. In Figures \ref{fig:AdamiW1}, \ref{fig:AdamiW3} and \ref{fig:AdamiW4} we show the detections for the common area of W1, W3 and W4. Generally, the large structures are well matched in all fields. In general, there is a better agreement with A10 2$\sigma$ detections. 

As a conclusion, a direct comparison of our work with A10 in data with similar depth shows a larger number of structures in our analysis, which are detected over 2$\sigma$  in the work by A10 but not over 3$\sigma$. In addition, there are some more structures that we find in this work and are not found in A10, which seem to be real (see section 5).

\begin{figure}
\centering
\includegraphics[clip,angle=90,scale=0.4]{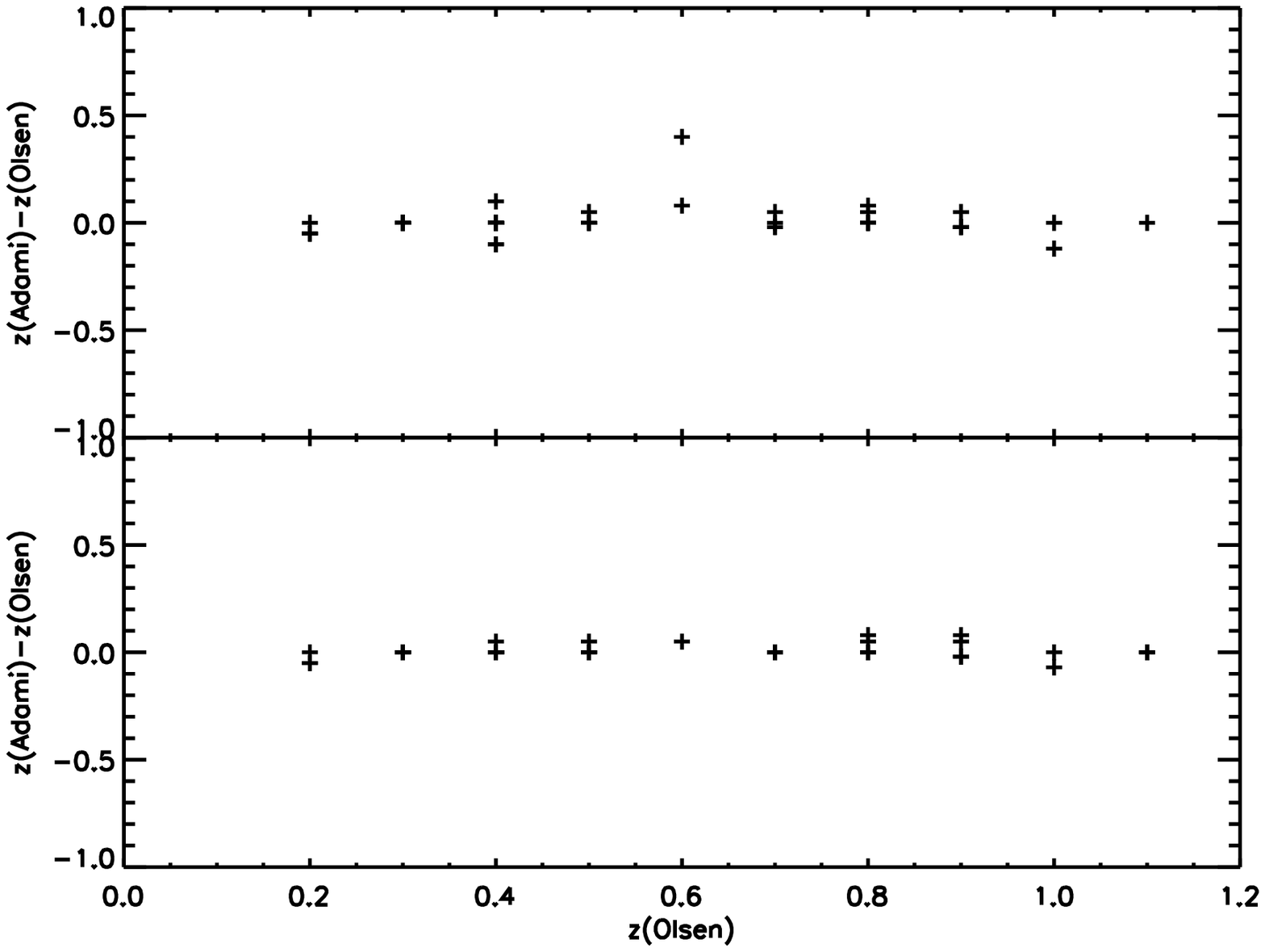} 
\caption{Redshift comparison between the detections from A10  over 3 and 2$\sigma$ and their matched detections by O07 (upper and bottom plot respectively).}
\label{fig:zOA}
\end{figure}

\begin{figure}
\centering
\includegraphics[clip,angle=90,scale=0.4]{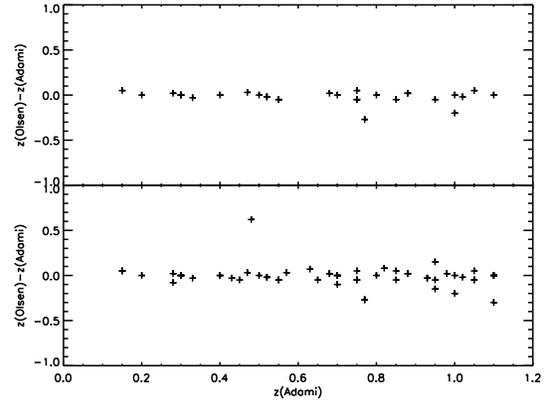} 
\caption{Redshift comparison between the O07 cluster and their matched detections from A10 in the area that overlaps with D1 over 3 and 2$\sigma$ (upper and bottom plot respectively).}
\label{fig:zAO}
\end{figure}

Eventually, we matched the detections from A10 to O07 catalogue using the same criteria. In Figure \ref{fig:zOA}, we show the redshift difference between the detections in O07 matched to the detections in A10 over 3 and 2$\sigma$ respectively, with respect to the O07 cluster redshift. We find a very good agreement for all redshifts. By matching the O07 catalogue to the A10 catalogue detections, we also find a good agreement as can be seen in Figure \ref{fig:zAO}. The dispersion increases at higher $z_{Adami}$.

\subsubsection{Comparison with X-ray data and spectroscopically confirmed sample}

We checked the agreement between the clusters that we detected in the optical, and X-ray detections found in these fields. \cite{pacaud07} published 29 spectroscopically confirmed clusters in the XMM-LSS survey, 17 of which are in W1. Additionally,  O07 made a compilation of 18 X-ray spectroscopically confirmed clusters in the same fields, of which 8 appear in the  \cite{pacaud07}  sample. 

Of these 25 clusters, we matched them to our sample by following the same criteria as before and we detected 23 within a minimum comoving distance of less than 4 Mpc distance. Even if this result is promising, the results might be biased since almost all the spectroscopy sample is selected to be in the high end of the mass function for clusters, making these results easier to be detected.

\begin{figure}
\centering
\includegraphics[clip,angle=90,scale=0.4]{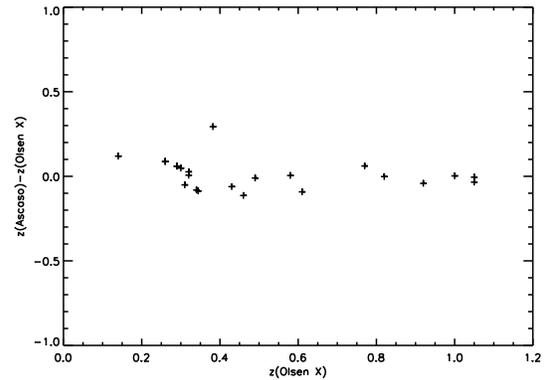} 
\caption{Redshift comparison between the X-ray spectroscopically detected clusters and our detections.}
\label{fig:zXrays}
\end{figure}

We find an excellent agreement between the redshift matches as we see  in Figure  \ref{fig:zXrays}. The median and dispersion of the redshift differences is 0.0024  and  0.088, well within the predicted errors.

\begin{figure}
\centering
\includegraphics[clip,angle=90,scale=0.4]{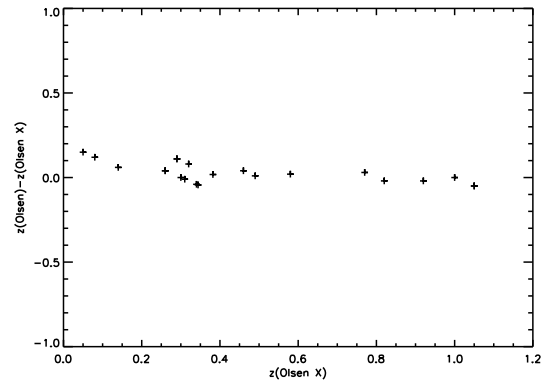} 
\caption{Redshift comparison between the X-ray spectroscopically detected clusters and O07 detections.}
\label{fig:zXraysOl}
\end{figure}

\begin{figure}
\centering
\includegraphics[clip,angle=90,scale=0.4]{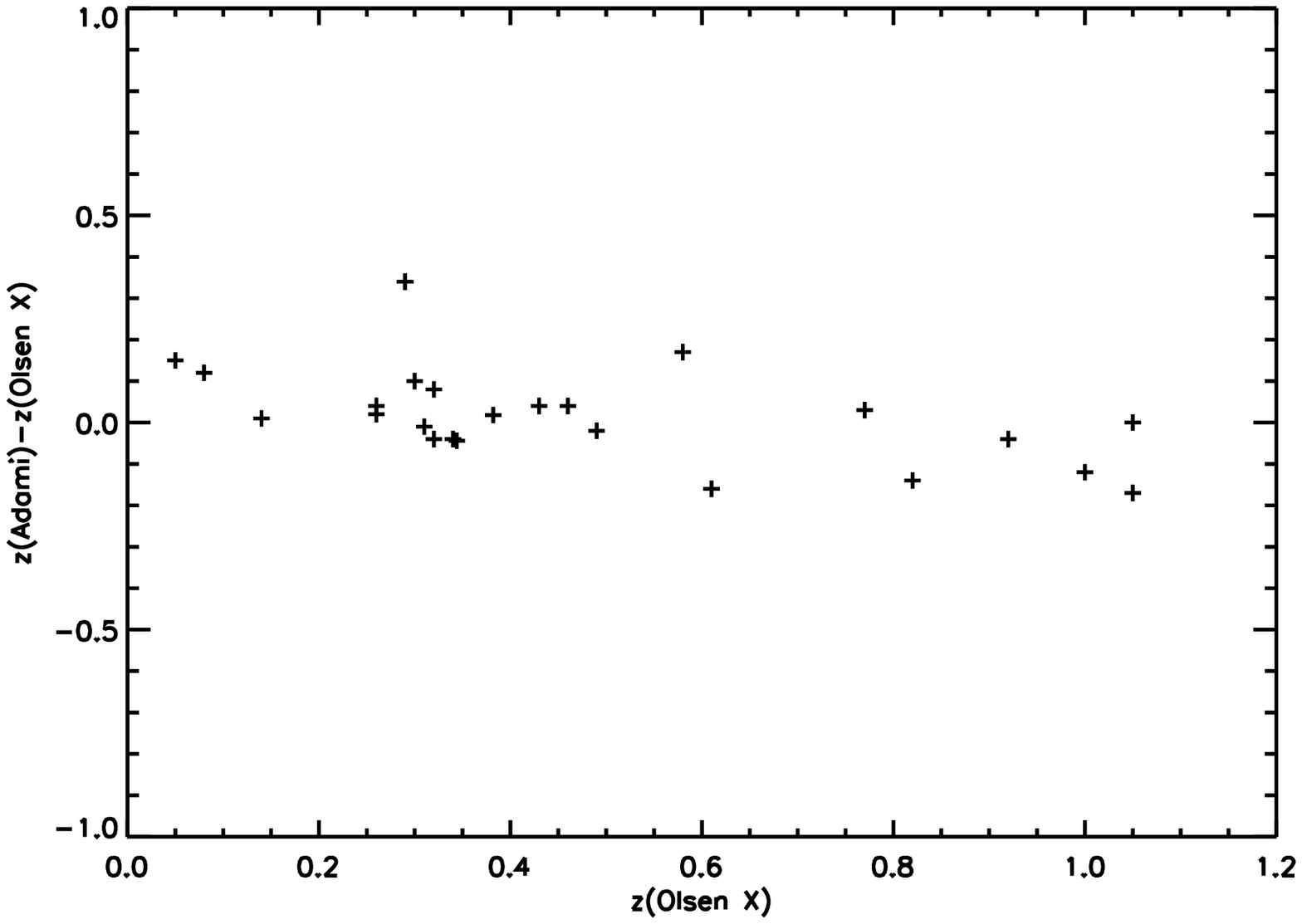} 
\caption{Redshift comparison between the X-ray spectroscopically detected clusters and A10 detections.}
\label{fig:zXraysAd}
\end{figure}

We also matched the sample of spectroscopically detection to O07 and A10 over 3$\sigma$. In Figures  \ref{fig:zXraysOl} and  \ref{fig:zXraysAd}, we show the redshift differences for the matched candidates to the sample. A10 finds all 25 galaxy clusters within 5 Mpc, where O07 finds 22. Let's note that A10's catalog is $\sim 1$ mag shallower than O07. However, the dispersion of the redshift difference matching is 0.058 and 0.1238 for A10 and O07 respectively.

\section{Discussion and conclusions}

We presented a new method based on a Bayesian approach of the matched filter technique, including the introduction of a prior containing the color-magnitude information, the BCG-redshift relation and the photometric redshift.  

We performed simulations to test how well our clusters are recovered. We simulated galaxy clusters showing red sequence at different redshift and with different richnesses. We found very high completeness rates ($>90\%$) and purity rates  for clusters with $\Lambda_{CL} \ge$ 20 ($>80\%$) up to redshift 1.2 at least. The results were very similar with and without the inclusion of the prior. This result suggests that galaxy clusters showing a red sequence can be found without the need to model the color-magnitude red sequence. In addition, the completeness and purity rates are high for a wide range of masses and redshifts.

It is still not clear if every galaxy cluster exhibits a red sequence. Different methods (X-ray, SZ, weak lensing) have been  finding these clusters independently of their colors in the high mass end and moderate redshift. Almost all of them appear to have a well defined red sequence. However, we do not know if this is the case at the low mass end of the cluster distribution or at high redshift. Hence, we tested the method on a \emph{mix} of clusters, ie: galaxy clusters showing and not showing a red sequence. The results could be different since when running the algorithm globally, we determined a background probability for each redshift slice for the whole field. This background probability is determined by the probability distribution of the galaxies that exist in the survey, including those galaxies in clusters. In this case, the galaxy clusters not showing a red sequence could affect the main background probability, making the non-red sequence clusters tail not to pass this threshold or, on the contrary, including spurious detections in the sample.

The results of completeness and purity for the whole mix of clusters are very similar to the red sequence galaxy cluster simulations, with a slight decrease of purity (but still $>80\%$) at higher redshift. As expected, the introduction of the prior in these simulations does not make a difference since those clusters are, by construction, not showing a red sequence. 
 
We applied this algorithm to the CARS Survey \citep{erben09} based on the CFHTLS Wide data and compared the results with the different works in the optical in the area corresponding to D1  by O07 and W1, W3 and W4 by A10. 

\begin{figure}
\centering
\includegraphics[clip,angle=90,scale=0.4]{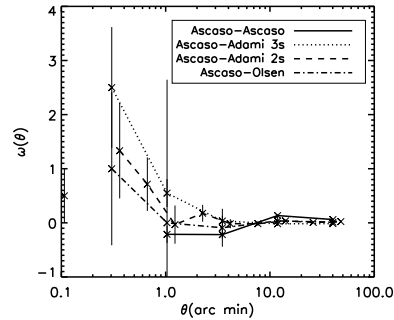} 
\caption{Cross-correlation as a function of angular separation of the cluster distribution in this work (solid line), between this work and A10 (dotted line; over 3$\sigma$, dashed line; over 2$\sigma$) and between this work and O07 (dashed-dotted line). We used the LS estimator.}
\label{fig:xcorrelation}
\end{figure}

It is hard to directly compare different works since every work uses a different merging and centering algorithm to obtain the final cluster catalogue. We calculated the cross-correlation between A10 and O07 and our work and it is shown in Figure \ref{fig:xcorrelation} by using the Landy-Szalay estimator (LS; \citealt{landy93}). We plot the auto correlation function as a reference. The cross-correlation function between two different sub-samples measures the excess probability over random of finding a cluster in the second sample at a given separation from a cluster belonging to the first sample. We find that both samples have a similar amplitude, being the clusters found by A10 over 3$\sigma$ the most correlated with the ones in this paper at smaller scales. 

\begin{figure}
\centering
\includegraphics[clip,angle=90,width=1.\hsize]{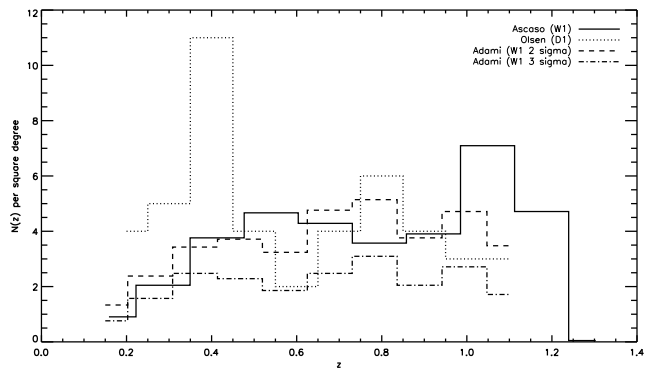} 
\caption{Redshift distribution for the clusters detections found in this work (solid line), O07 (dotted line) and A10 over 2 and 3$\sigma$ (dashed and dashed-dotted lines).}
\label{fig:zcompare}
\end{figure}

\begin{figure}
\centering
\includegraphics[clip,angle=0,scale=0.2]{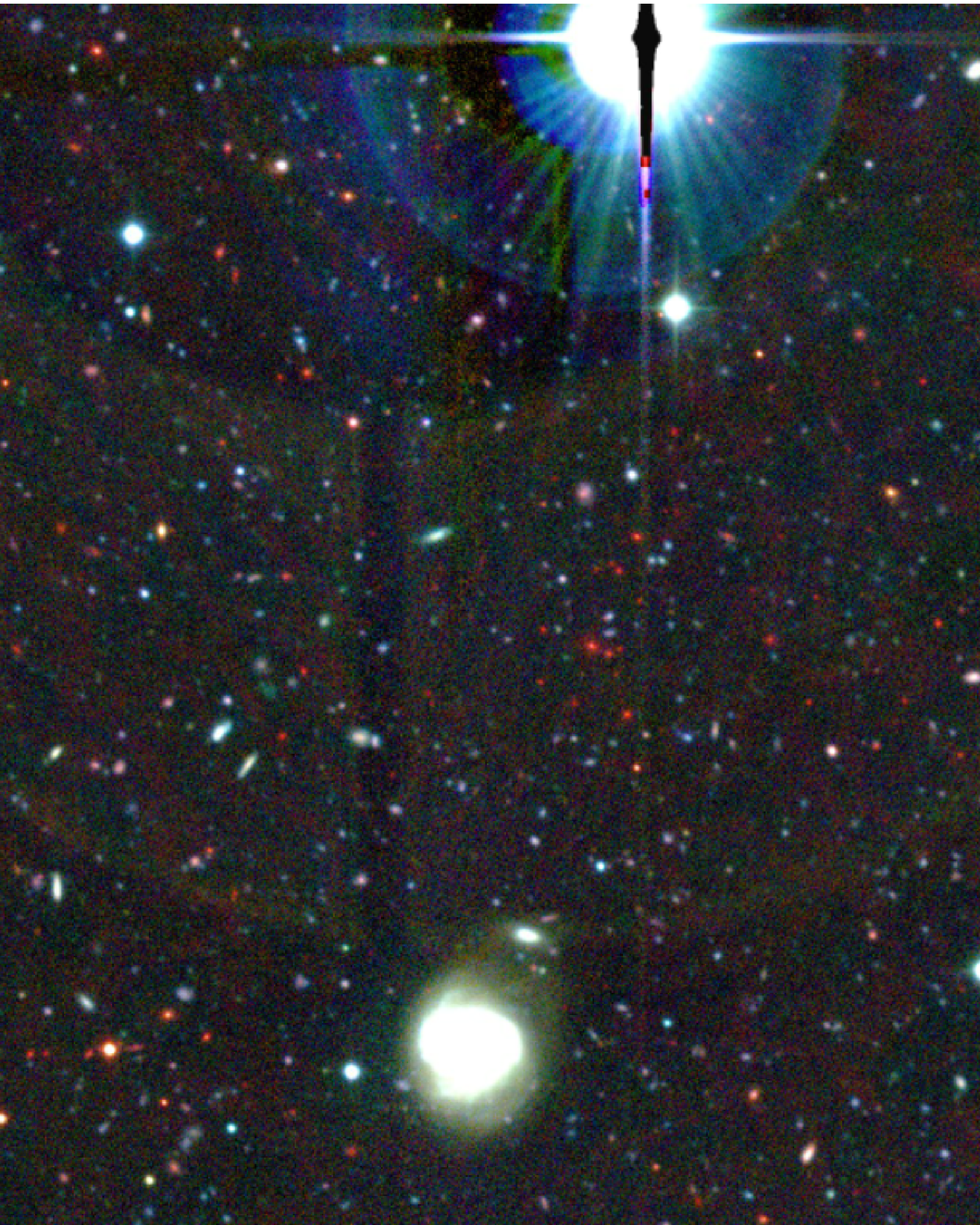} 
\includegraphics[clip,angle=0,scale=0.25]{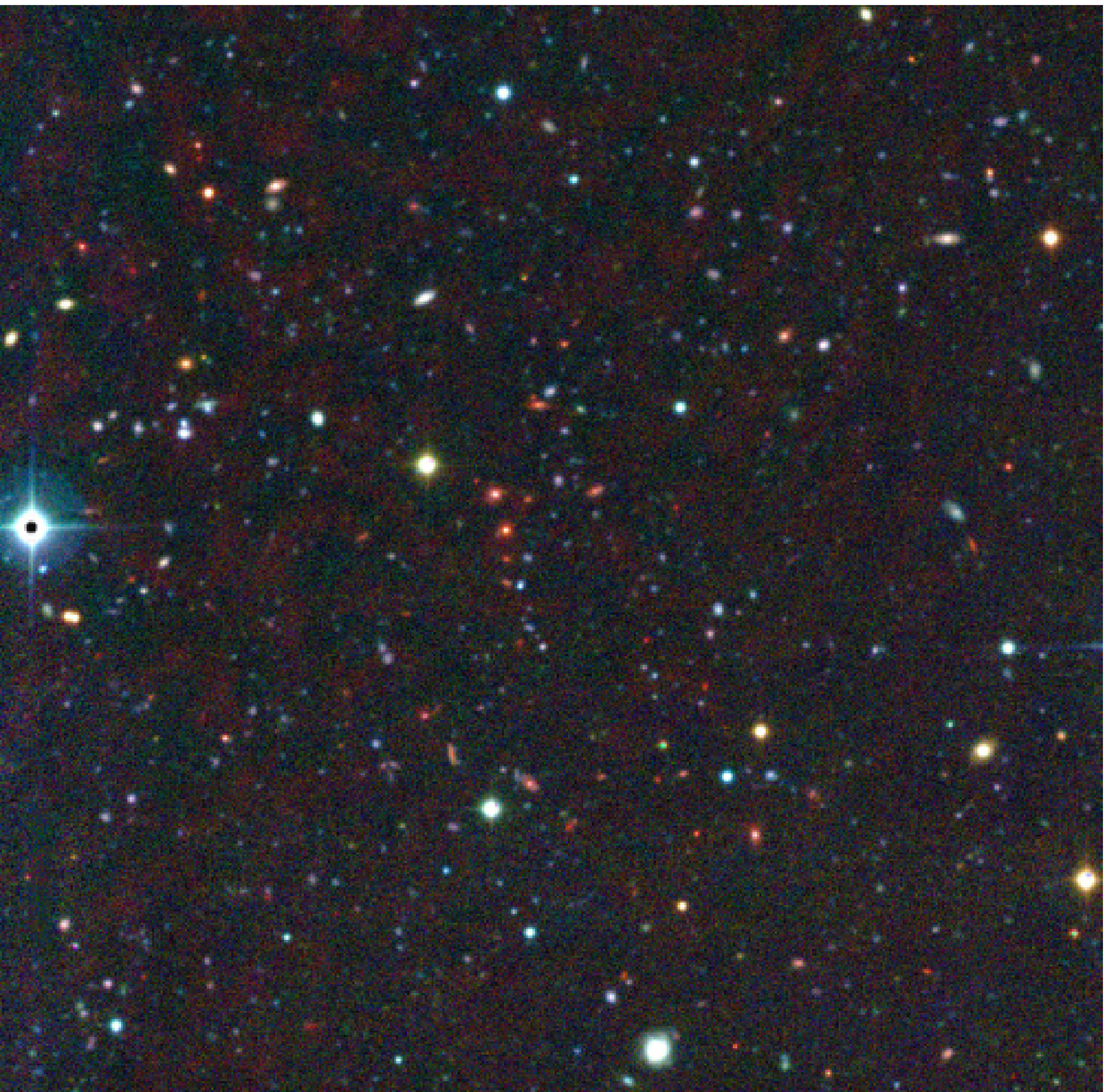} 
\includegraphics[clip,angle=0,scale=0.2]{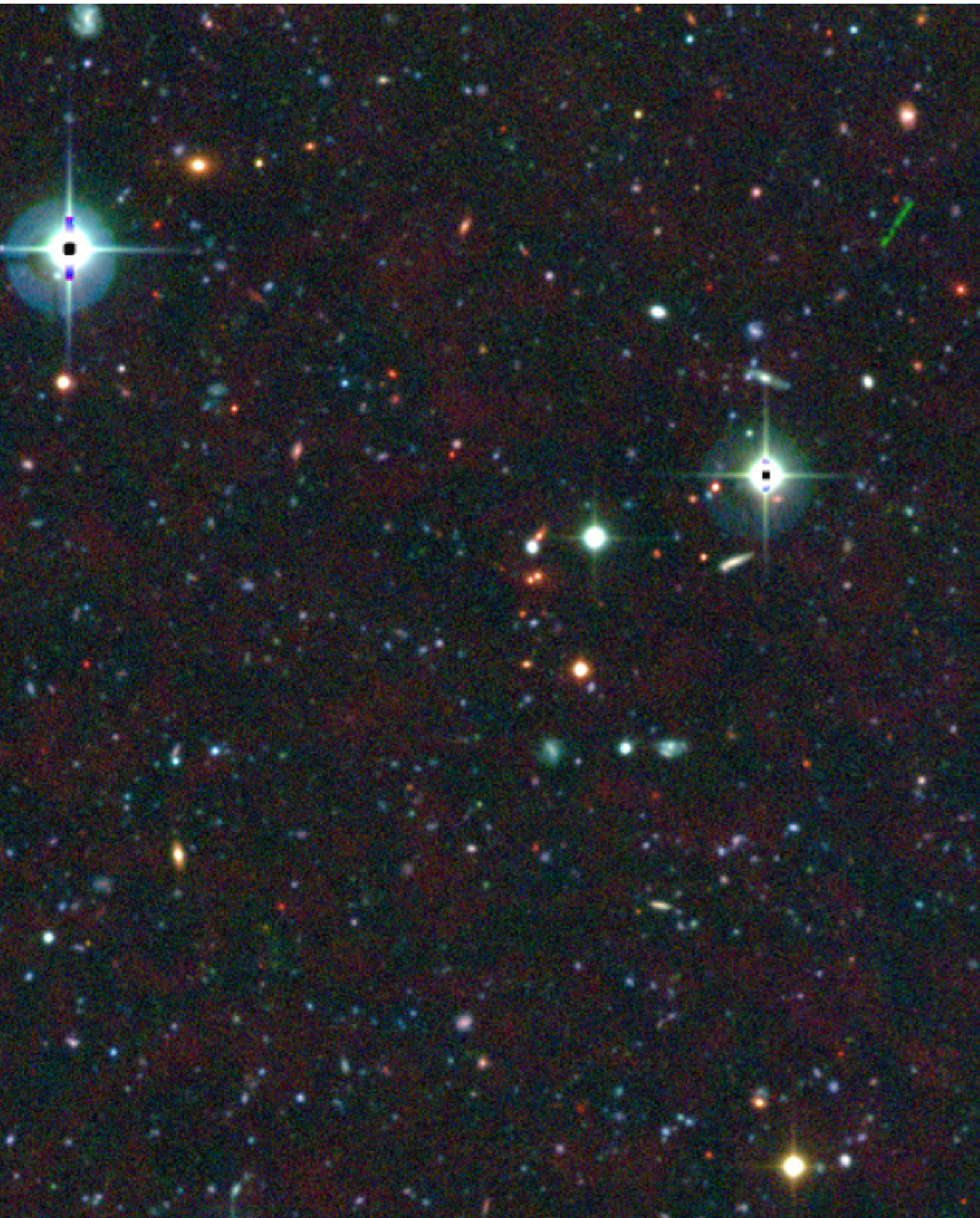} 
\caption{Color images of a galaxy cluster detections found by this work in the common area of CARS with Deep 1. The works by O07 and A10 do not detect them. However, the structures seem to be real since we find a concentration or similar color galaxies. The upper, middle and bottom detections have estimated redshifts of 1.01, 0.92 and 1.21 respectively. In fact, the upper structure was detected with X-rays in Pacaud et al. 2007 with a redshift estimate of 1.05. The size of the box corresponds to a $\sim$ 1 Mpc size box at the estimated redshift of the structure.}
\label{fig:clusters}
\end{figure}

In Figure  \ref{fig:zcompare}, we show the redshift distribution of the detections found by O07, A10 and this work. O07 finds a peak of detections at z $\sim$ 0.4, which are mostly low mass structures detected in their deeper data. Apart from this, the redshift detection distribution in this works agrees  well with the redshift distribution from O07 and with the detections found by  A10 over  2$\sigma$ better than 3$\sigma$.  However, we find a peak of detections at z  $\sim$ 1 that the other works do not find. Many of these cluster candidates seem to be real by looking at the images as shown in Figure \ref{fig:clusters}. In fact, some of them were detected with X-rays in \cite{pacaud07} as the first cluster in Figure  \ref{fig:clusters}. They provided a redshift estimate of 1.05 for this cluster, whereas we obtain a redshift estimate of 1.01. These detections will be confirmed by obtaining spectroscopic information or deeper data in these regions.

In summary, we find a good agreement with the galaxy clusters that we find in CARS and other detections found in similar or deeper data of the same fields. In addition, we find several detections at high redshift which are not reported and appear to be real based on visual characteristics. We are currently applying this algorithm to the Deep Lens Survey (DLS), (Ascaso et al. in preparation) in order to provide the first optical galaxy cluster in the survey.

\section*{Acknowledgments}
We are thankful to the anonymous referee who improved the quality of this paper with useful suggestions. We acknowledge Thomas Erben for making available their catalogs and data for the CARS survey. Narciso Ben\'itez acknowledges support from AYA2010-22111-C03-00. Bego\~na Ascaso thanks Lori Lubin, Vera Margoniner, Will Dawson and Tony Tyson for helpful discussion. We acknowledge the DLS group for valuable suggestions to this paper.


\begin{thebibliography}{99}
\bibitem[Adami et al.(2010)]{adami10} Adami, C., et al.\ 2010, \aap, 509, A81
\bibitem[Andreon et al.(2008)]{andreon08} Andreon, S., Maughan, B., Trinchieri, G., \& Kurk, J.\ 2008, arXiv:0812.169
\bibitem[Ascaso \& Moles(2007)]{ascaso07} Ascaso, B., \& Moles, M.\ 2007, \apjl, 660, L89
\bibitem[Ascaso et al.(2008)]{ascaso08} Ascaso, B., Moles, M., Aguerri, J.~A.~L., S{\'a}nchez-Janssen, R., \& Varela, J.\ 2008, \aap, 487, 453
\bibitem[Ascaso et al.(2011)]{ascaso11} Ascaso, B., Aguerri, J.~A.~L., Varela, J., Cava, A., Bettoni, D., Moles, M.,  \& D'Onofrio, M.\ 2011, \apj, 726, 69
\bibitem[Baldry et al.(2004)]{baldry04} Baldry, I.~K., Glazebrook, K., Brinkmann, J., et al. \ 2004, \apj, 600, 681 
\bibitem[Ben{\'{\i}}tez(2000)]{benitez00} Ben{\'{\i}}tez, N.\ 2000, \apj, 536, 571 
\bibitem[Blanton et al.(2003)]{blanton03} Blanton, M.~R., et al.\ 2003, \apj, 592, 819
\bibitem[Blanton \& Roweis(2007)]{blanton07} Blanton, M.~R., \& Roweis, S.\ 2007, \aj, 133, 734 
\bibitem[Botzler et al.(2004)]{botzler04} Botzler, C.~S., Snigula, J., Bender, R., \& Hopp, U.\ 2004, \mnras, 349, 425 
\bibitem[Brown et al.(2007)]{brown07} Brown, M.~J.~I., Dey, A., Jannuzi, B.~T., Brand, K., Benson, A.~J., Brodwin, M., Croton, D.~J., 
\& Eisenhardt, P.~R.\ 2007, \apj, 654, 858 
\bibitem[Carlstrom et al.(2002)]{carlstrom02} Carlstrom, J.~E., Holder, G.~P., \& Reese, E.~D.\ 2002, \araa, 40, 643 
\bibitem[Chiaberge et al.(2010)]{chiaberge10} Chiaberge, M., Capetti, A., Macchetto, F.~D., Rosati, P., Tozzi, P., \& Tremblay, G.~R.\ 2010, \apjl, 710, L107 
\bibitem[Coleman et al.(1980)]{coleman80} Coleman, G.~D., Wu, C.-C., \& Weedman, D.~W.\ 1980, \apjs, 43, 393
\bibitem[Couch et al.(1991)]{couch91} Couch, W.~J., Ellis, R.~S., MacLaren, I., \& Malin, D.~F.\ 1991, \mnras, 249, 606
\bibitem[Dalton et al.(1997)]{dalton97} Dalton, G.~B., Maddox, S.~J., Sutherland, W.~J., \& Efstathiou, G.\ 1997, \mnras, 289, 263
\bibitem[Demarco et al.(2010)]{demarco10} Demarco, R., et al.\ 2010, \apj, 711, 1185 
\bibitem[Donahue et al.(2002)]{donahue02} Donahue, M., et al.\ 2002, \apj, 569, 689  
\bibitem[Dong et al.(2008)]{dong08} Dong, F., Pierpaoli, E., Gunn, J.~E., \& Wechsler, R.~H.\ 2008, \apj, 676, 868
\bibitem[Eisenhardt et al.(2008)]{eisenhardt08} Eisenhardt, P.~R.~M., et al.\ 2008, \apj, 684, 905 
\bibitem[Erben et al.(2009)]{erben09} Erben, T., et al.\ 2009, \aap, 493, 1197
\bibitem[Gal et al.(2000)]{gal00} Gal, R.~R., de Carvalho, R.~R., Odewahn, S.~C., Djorgovski, S.~G., \& Margoniner, V.~E.\ 2000, \aj, 119, 12
\bibitem[Gal et al.(2003)]{gal03} Gal, R.~R., de Carvalho, R.~R., Lopes, P.~A.~A., Djorgovski, S.~G., Brunner, R.~J., Mahabal, A., \& Odewahn, S.~C.\ 2003, \aj, 125, 2064
\bibitem[Galametz et al.(2009)]{galametz09} Galametz, A., et al.\ 2009, \aap, 507, 131 
\bibitem[Gladders \& Yee(2000)]{gladders00} Gladders, M.~D., \& Yee, H.~K.~C.\ 2000, \aj, 120, 2148 
\bibitem[Gladders \& Yee(2005)]{gladders05} Gladders, M.~D., \& Yee, H.~K.~C.\ 2005, \apjs, 157, 1
\bibitem[Grazian et al.(2006)]{grazian06} Grazian, A., et al.\ 2006, \aap, 453, 507
\bibitem[Grove et al.(2009)]{grove09} Grove, L.~F., Benoist, C., \& Martel, F.\ 2009, \aap, 494, 845
\bibitem[Goto et al.(2002)]{goto02} Goto, T., et al.\ 2002, \aj, 123, 1807
\bibitem[Hansen et al.(2005)]{hansen05} Hansen, S.~M., McKay, T.~A., Wechsler, R.~H., Annis, J., Sheldon, E.~S., \& Kimball, A.\ 2005, \apj, 633, 122 
\bibitem[Harsono \& DePropris(2009)]{harsono09} Harsono, D., \& DePropris, R.\ 2009, \aj, 137, 3091
\bibitem[Huchra \& Geller(1982)]{huchra82} Huchra, J.~P., \& Geller, M.~J.\ 1982, \apj, 257, 423
\bibitem[Ilbert et al.(2006)]{ilbert06} Ilbert, O., et al.\ 2006, \aap, 457, 841 
\bibitem[Kepner et al.(1999)]{kepner99} Kepner, J., Fan, X., Bahcall, N., Gunn, J., Lupton, R., \& Xu, G.\ 1999, \apj, 517, 78 
\bibitem[Kim et al.(2002)]{kim02} Kim, R.~S.~J., et al.\ 2002, \aj, 123, 20 
\bibitem[Kinney et al.(1996)]{kinney96} Kinney, A.~L., et al.\ 1996, \apj, 467, 38 
\bibitem[Koester et al.(2007)]{koester07} Koester, B.~P., et al.\ 2007, \apj, 660, 221 
\bibitem[Koester et al.(2007)]{koester07b} Koester, B.~P., et al.\ 2007, \apj, 660, 239
\bibitem[Landy \& Szalay(1993)]{landy93} Landy, S.~D., \& Szalay, A.~S.\ 1993, \apj, 412, 64 
\bibitem[Lidman \& Peterson(1996)]{lidman96} Lidman, C.~E., \& Peterson, B.~A.\ 1996, \aj, 112, 2454 
\bibitem[Lopes et al.(2004)]{lopes04} Lopes, P.~A.~A., de Carvalho, R.~R., Gal, R.~R., Djorgovski, S.~G., Odewahn, S.~C., Mahabal, A.~A., \& Brunner, R.~J.\ 2004, \aj, 128, 1017 
\bibitem[L{\'o}pez-Cruz et al.(2004)]{lopezcruz04} L{\'o}pez-Cruz, O., Barkhouse, W.~A., \& Yee, H.~K.~C.\ 2004, \apj, 614, 679
\bibitem[Mei et al.(2006)]{mei06} Mei, S., et al.\ 2006, \apj, 644, 759 
\bibitem[Mei et al.(2009)]{mei09} Mei, S., et al.\ 2009, \apj, 690, 42 
\bibitem[Menanteau et al.(2009)]{menanteau09} Menanteau, F., et al.\ 2009, \apj, 698, 1221 
\bibitem[Miller et al.(2005)]{miller05} Miller, C.~J., et al.\ 2005, \aj, 130, 968 
\bibitem[Milkeraitis et al.(2010)]{milkeraitis10} Milkeraitis, M., van Waerbeke, L., Heymans, C., Hildebrandt, H., Dietrich, J.~P., \& Erben, T.\ 2010, \mnras, 701 
\bibitem[Moran et al.(2007)]{moran07} Moran, S.~M., Ellis, R.~S., Treu, T., Smith, G.~P., Rich, R.~M., \& Smail, I.\ 2007, \apj, 671, 1503 
\bibitem[Muzzin et al.(2009)]{muzzin09} Muzzin, A., et al.\ 2009, \apj, 698, 1934 
\bibitem[Nakata et al.(2001)]{nakata01} Nakata, F., et al.\ 2001, \pasj, 53, 1139
\bibitem[Nakamura et al.(2003)]{nakamura03} Nakamura, O. et al. \ 2003, \aj, 125, 1682 
\bibitem[Navarro et al.(1997)]{navarro97} Navarro, J.~F., Frenk, C.~S., \& White, S.~D.~M.\ 1997, \apj, 490, 493 
\bibitem[Olsen et al.(2007)]{olsen07} Olsen, L.~F., et al.\ 2007, \aap, 461, 81
\bibitem[Pacaud et al.(2007)]{pacaud07} Pacaud, F., et al.\ 2007, \mnras, 382, 1289 
\bibitem[Postman et al.(1996)]{postman96} Postman, M., Lubin, L.~M., Gunn, J.~E., Oke, J.~B., Hoessel, J.~G., Schneider, D.~P., \& Christensen, J.~A.\ 1996, \aj, 111, 615 
\bibitem[Postman et al.(2001)]{postman01} Postman, M., Lubin, L.~M., \& Oke, J.~B.\ 2001, \aj, 122, 1125
\bibitem[Postman et al.(2002)]{postman02} Postman, M., Lauer, T.~R., Oegerle, W., \& Donahue, M.\ 2002, \apj, 579, 93
\bibitem[Ramella et al.(2001)]{ramella01} Ramella, M., Boschin, W., Fadda, D., \& Nonino, M.\ 2001, \aap, 368, 776
\bibitem[Ramella et al.(2002)]{ramella02} Ramella, M., Geller, M.~J., Pisani, A., \& da Costa, L.~N.\ 2002, \aj, 123, 2976
\bibitem[Rosati et al.(2002)]{rosati02} Rosati, P., Borgani, S., \& Norman, C.\ 2002, \araa, 40, 539
\bibitem[Schechter(1976)]{schechter76} Schechter, P.\ 1976, \apj, 203, 297 
\bibitem[Shectman(1985)]{shectman85} Shectman, S.~A.\ 1985, \apjs, 57, 77 
\bibitem[Schuecker \& Boehringer(1998)]{schuecker98} Schuecker, P., \& Boehringer, H.\ 1998, \aap, 339, 315
\bibitem[Szabo et al.(2010)]{szabo10} Szabo, T., Pierpaoli, E., Dong, F., Pipino, A., \& Gunn, J.~E.\ 2010, arXiv:1011.0249
\bibitem[Treu et al.(2003)]{treu03} Treu, T., Ellis, R.~S., Kneib, J.-P., Dressler, A., Smail, I., Czoske, O., Oemler, A., \& Natarajan, P.\ 2003, \apj, 591, 53 
\bibitem[Tyson et al.(1990)]{tyson90} Tyson, J.~A., Wenk, R.~A., \& Valdes, F.\ 1990, \apjl, 349, L1 
\bibitem[van Breukelen \& Clewley(2009)]{vanbreukelen09} van Breukelen, C., \& Clewley, L.\ 2009, \mnras, 395, 1845 
\bibitem[Way et al.(2010)]{way10} Way, M.~J., Gazis, P.~R., \& Scargle, J.~D.\ 2010, arXiv:1009.0387 
\bibitem[White \& Frenk(1991)]{white91} White, S.~D.~M., \& Frenk, C.~S.\ 1991, \apj, 379, 52
\bibitem[Wilson et al.(2006)]{wilson06} Wilson, G., Muzzin, A., Lacy, M., \& FLS Survey Team 2006, Astronomical Society of the Pacific Conference Series, 357, 238 
\bibitem[Wilson et al.(2008)]{wilson08} Wilson, G., et al.\ 2008, Infrared Diagnostics of Galaxy Evolution, 381, 210 
\bibitem[Wilson et al.(2009)]{wilson09} Wilson, G., et al.\ 2009, \apj, 698, 1943 
\bibitem[Wittman et al.(2001)]{wittman01} Wittman, D., Tyson, J.~A., Margoniner, V.~E., Cohen, J.~G., \& Dell'Antonio, I.~P.\ 2001, \apjl, 557, L89
\bibitem[Wittman et al.(2003)]{wittman03} Wittman, D., Margoniner, V.~E., Tyson, J.~A., Cohen, J.~G., Becker, A.~C., \& Dell'Antonio, I.~P.\ 2003, \apj, 597, 218 
\bibitem[Zaritsky et al.(1997)]{zaritsky97} Zaritsky, D., Nelson, A.~E., Dalcanton, J.~J., \& Gonzalez, A.~H.\ 1997, \apjl, 480, L91
\bibitem[Zaritsky et al.(2002)]{zaritsky02} Zaritsky, D., Gonzalez, A.~H., Nelson, A.~E., \& Dalcanton, J.~J.\ 2002, AMiBA 2001: High-Z Clusters, Missing Baryons, and CMB Polarization, 257, 133
\end{thebibliography}
\end{document}